\documentclass[prb,twocolumn,superscriptaddress,showpacs,floatfix]{revtex4}
\usepackage{lipsum}
\usepackage{amsmath,amssymb,bm,bbm} 
\usepackage{graphicx}  
\usepackage{wasysym}
\usepackage{color} 
\usepackage{relsize} 
\usepackage[utf8]{inputenc}
\usepackage{bbm}
\usepackage{mathrsfs}
\usepackage{epsfig,psfrag}
\usepackage[final]{pdfpages}
\usepackage{pstool}
\usepackage{anyfontsize}
\usepackage{rotating}
\usepackage{longtable}
\usepackage{braket}
\def\be{\begin{equation}}
\def\ee{\end{equation}}
\def\bea{\begin{eqnarray}}
\def\eea{\end{eqnarray}}

\def\Minus{\texttt{-}}
\begin{document}
\title{Exotic topological point and line nodes in the plaquette excitations of a  
frustrated Heisenberg antiferromagnet on the honeycomb lattice}
\author{Moumita Deb}\email{moumitadeb44@gmail.com}
\author{Asim Kumar Ghosh}
 \email{asimkumar96@yahoo.com}
\affiliation {Department of Physics, Jadavpur University, 
188 Raja Subodh Chandra Mallik Road, Kolkata 700032, India}
\begin{abstract}
A number of topological nodes including Dirac, quadratic and three-band 
touching points as well as a pair of degenerate Dirac line nodes are found to emerge in the 
triplet plaquette excitations 
of the frustrated spin-1/2 $J_1$-$J_2$ antiferromagnetic Heisenberg honeycomb model 
when the ground state of the system lies in a spin-disordered plaquette-valence-bond-solid phase. 
A six-spin plaquette operator theory of this honeycomb model 
has been developed for this purpose by using the eigenstates of an isolated 
Heisenberg hexagonal plaquette. 
Spin-1/2 operators are thus expressed in the Fock space spanned by the plaquette 
operators those are obtained in terms of exact analytic form of eigenstates for a 
single frustrated Heisenberg hexagon. 
Ultimately, an effective interacting boson model of this system is obtained on the 
basis of low energy singlets and triplets plaquette operators by employing 
a mean-field approximation.  
The values of ground state energy and spin gap of this system have been estimated 
and the validity of this formalism has been tested upon 
comparison with the known results. 
Emergence of topological point and line nodes on the basis of spin-disordered ground 
state noted in this investigation is very rare on any frustrated system 
as well as the presence of triplet flat band.   
Evolution of those topological nodes is studied throughout the full 
frustrated regime. Finally, emergence of topological phases 
has been reported upon adding a time-reversal-symmetry 
breaking term to the Hamiltonian.
Coexistence of spin gap with either topological nodes or
phases turns this honeycomb model an interesting one. 
\end{abstract}

\maketitle

\section{INTRODUCTION}
The observation of magnon Hall effect in ferromagnetic (FM) 
compound Lu$_2$V$_2$O$_7$ has given further impetus toward the 
investigations of topological phases in magnetic systems \cite{Tokura}. 
Search of topological phases 
in the antiferromegnetic (AFM) systems begins afterwards 
as a consequence of this observation. 
Most of the studies involve in finding nontrivial topology 
in magnon bands on the basis of spin-ordered ground states. 
The frustrated AFM systems, on the other hand, very often give rise to 
exotic spin-disordered ground states those are generally studied 
in terms of either valence-bond-solid (VBS) or resonating-valence-bond 
(RVB) states \cite{Sachdev1,Misguich}.  
Thus, search of topological phases nowadays extended 
beyond the magnetic systems of having spin-ordered ground states. 
But the finding of topological nontriviality 
based on the spin-disordered ground state is more challenging 
for several reasons. In this investigation,  
emergence of topological point and line nodes 
along with nontrivial topological phases will be 
reported in a frustrated AFM spin-1/2 Heisenberg model 
on the honeycomb lattice where the ground state is a 
plaquette-VBS (PVBS) state.

In order to investigate the dynamics of the 
AFM Heisenberg model on the honeycomb lattice 
a six-spin plaquette operator theory (POT) 
in terms of triplet plaquette excitations 
has been developed on the basis of singlet 
PVBS ground state. Several theoretical approaches have been employed 
before to study the ground state properties and the
dynamics of the AFM $J_1$-$J_2$ model on honeycomb lattice, where, 
$J_1$ and $J_2$ are the nearest (NN) and next-nearest (NNN) neighbor 
exchange strengths, respectively \cite{Albuquerque,Baskaran,
Meng,Sondhi,Jafari,Lamas,Bishop,Ganesh,Zhu,Gong,Oitmaa,Fouet,Becca}. 
Most of the studies limit 
themselves within the moderate frustration range, $0.0<J_2/J_1<0.5$.  
Existence of three distinct quantum phases has been marked in this regime.  
One among them is a spin-disordered phase which lies in 
the intermediate regime, $0.2<J_2/J_1<0.4$, 
between two different ordered regimes. 
The ordered phases are 
gapless N\'eel and spiral states, which survive in the regions, 
$0.0<J_2/J_1<0.2$ and $0.4<J_2/J_1<0.5$, respectively. 
Previous studies have taken 
this phase diagram for granted with a little dispute on the location of 
boundaries separating the different phases. 
The nature of disordered state in the intermediate region 
is not free of ambiguity as well. But most of the recent studies support   
the existence of PVBS state \cite{Albuquerque,Becca}. 
The signature of PVBS phase has been detected in the AFM honeycomb 
compound LiZn$_2$Mo$_3$O$_8$ by measuring the triplet spin gap \cite{Lee}. 
On the other hand, AFM spin ordered phase has been detected in Na$_2$IrO$_3$, 
and the thermodynamic properties of this compound have been explained 
in terms of $J_1$-$J_2$ Heisenberg honeycomb model with 
$J_2/J_1=0.47$ \cite{Singh}. 

In this investigation, POT has been formulated for the entire 
frustrated regime $0<J_2/J_1<1$ of the model based on two orthogonal 
plaquette-RVB (PRVB) states, separately for the 
moderate ($0<J_2/J_1<1/2$) and extreme ($1/2<J_2/J_1<1$) frustrated regimes. 
Those PRVB states are not only the exact ground states of a single hexagonal plaquette, 
separately for the two different frustrated regimes but also singlet.  
Dynamics of the system are studied in terms of a low-energy mean-field 
Hamiltonian, where dispersion relations of three lowest energy triplets are obtained. 
Ground state energy and spin gap have been estimated and compared with the 
numerical results. 

Surprisingly, examining the bosonic triplet dispersion relation 
of this honeycomb model, several types of topological point and line nodes 
are found to emerge upon the variation of $J_2/J_1$. 
Two-band and three-band touching points with two different 
kinds for each one have been noted. 
Two-band touching nodes are identified as either 
Dirac or quadratic band touching points (QBTP) \cite{Soljacic}. 
Similarly, the three-band touching points in the two regimes 
are qualitatively different which will be discussed later. 

In addition, a flat band and a pair of degenerate 
Dirac line nodes (DLN) 
are found in the extreme frustrated regime. 
DLN is formed when two linear dispersions touch over a line 
on the Brillouin zone (BZ) rather than at a point. 
Both the DLNs are topologically protected by the simultaneous existence of 
space-inversion and time-reversal symmetry (${\cal PT}$-symmetry) 
of the system \cite{Fang,Mertig}. 
One Dirac node among all of them is 
protected by the symmetry of the system, since its position 
in the BZ is fixed regardless the values of $J_2/J_1$, 
which is analogous to that observed in graphene \cite{Wallace}. 
While the remaining nodes are 
tunable in a sense that their characteristics can be 
changed by varying the parameters. 
Thus this frustrated honeycomb model hosts a variety of 
topological nodes which are not seen before in a single model.  

No topological node is found in the bosonic magnon excitation of the 
AFM Heisenberg model on the honeycomb lattice, though, emergence of a solitary 
Dirac node is reported in the magnon excitation 
of a FM Heisenberg model \cite{Boyko,Owerre1}. 
Additional Weyl nodes emerge when next-next-nearest-neighbor (NNNN) 
interactions are taken into account in the FM case \cite{Boyko}. 
The magnon DLN is found before in FM Heisenberg model on the 
three-dimensional (3D) pyrochlore lattice and AFM Heisenberg model 
on the 2D square-octagon lattice \cite{Mertig,Owerre2}. 
However, all these topological nodes reported before 
are based on the spin-ordered ground states. 
In contrast, all the nodal points and lines emerged 
in this investigation are based on the spin-disordered ground states. 

Chern number (C) acts as an invariant for a particular 
class of topological phases when the 
time-reversal symmetry (${\cal T}$-symmetry) of 
the system is broken \cite{TKNN}. 
Topological protection of the insulating bulk bands is 
additionally rewarded by the presence of in-gap edge states connecting the 
separated bands. The value of C and the number of edge state 
modes are related by the `bulk-edge-correspondence' (BEC) rule \cite{Hatsugai}. 
Previous studies reveal that 
a gap in the magnon excitation 
of the FM Heisenberg model opens up at the Dirac nodal point 
as soon as the NNN Dzyaloshinskii-Moriya interaction (DMI) is invoked,  
where DMI breaks the space-inversion symmetry 
(${\cal P}$-symmetry) \cite{Owerre1}.
At the same time, the system becomes topologically nontrivial 
with C=$\pm 1$. Observation of this particular topological feature 
has been claimed in the FM honeycomb compound CrI$_3$
by examining the magnon band obtained in inelastic neutron scattering \cite{Chen}. 
FM Heisenberg models on the honeycomb lattice 
with Kitaev and spin-anisotropic interactions (SAI) 
are found to host a number of topological phases \cite{Joshi,Deb1}. 
Here, the Zeeman term corresponding to the external 
magnetic field breaks the ${\cal T}$-symmetry. 
However, in these cases, topological phases 
are found in those excitation bands which are based on the spin-ordered 
ground states. Also, no topological phase 
based on AFM spin-ordered ground state is reported on the 
honeycomb model. 
Motivated by the emergence of multiple topological nodes of 
various kinds in this system, the search of Chern insulating phases in the 
triplet excitation bands based on the disordered ground state  
has been undertaken in this investigation. 
The system in the moderate frustrated region 
is found to host six distinct topological 
phases when the effective three-band 
Hamiltonian losses its ${\cal T}$-symmetry invariance. 

Two-spin bond operator theory 
was introduced before by Sachdev and
Bhat to study the ground state phase diagram of an AFM Heisenberg model on
the square lattice in terms of several VBS states 
on the basis of various singlet dimer coverings \cite{Sachdev2}. 
The method has been extended up to four-spin plaquette operator 
to study the properties of frustrated AFM  
Heisenberg models on square-octagon and square lattices 
based on the PVBS ground states \cite{Ueda,Doretto}. 
Upon further extension, POT has 
been developed on the basis of 
six-spin frustrated AFM Heisenberg plaquette in this investigation. 

The article has been organized in the following way. 
Properties of a single frustrated Heisenberg hexagonal plaquette is 
described in the Section \ref{SHP}. POT has 
been described in the Section \ref{POT}. 
An effective low-energy Hamiltonian 
in terms of bosonic singlets and triplets operators for $J_1$-$J_2$
AFM Heisenberg model on the honeycomb lattice has been 
formulated in the Section \ref{LEBM}. 
To estimate the ground state energy and spin gap of the frustrated 
system a mean-field theory has been developed in the 
Section \ref{MFA}. Emergence and evolution of 
topological nodal points and lines 
are described in the Section \ref{TN} and the properties of 
topological phases are explained in Section \ref{TP}. 
Section \ref{discussion} holds a comprehensive discussion on the
results.

 \begin{figure*}[t]
  \includegraphics [width=460pt]{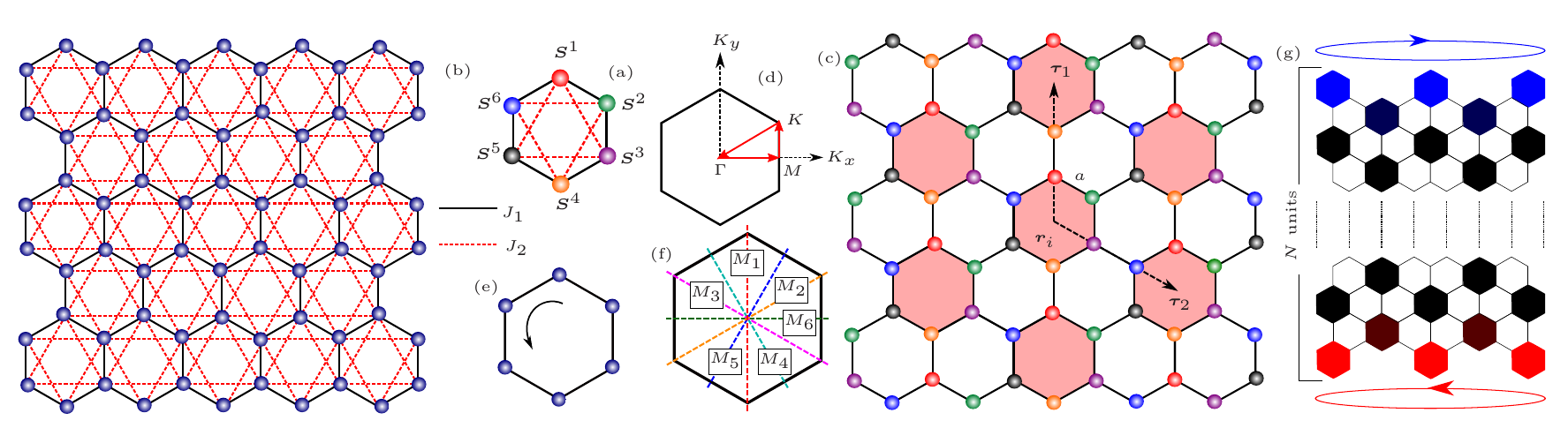}
    \caption{(a) Geometrical view of the hexagonal plaquette, (b) $J_1$-$J_2$ AFM Heisenberg honeycomb lattice,  
(c) honeycomb lattice of plaquettes, (d) Brillouin zone of the
plaquette honeycomb lattice defined by $\Gamma=(0,0)$, $M=(\frac{2\pi}{3\sqrt{3}},0)$ and $K=(\frac{2\pi}{3\sqrt{3}},\frac{2\pi}{9})$, 
(e) system remains invariant under rotation by $\pi/3$ ,
(f) mirror planes are shown by six different dashed lines, 
(g) strip of honeycomb lattice used
for edge state calculation, upper and lower edges are drawn in blue and red colors, respectively.}
  \label{lattice}
  \end{figure*} 

\section{Single hexagonal plaquette}
\label{SHP}
The spin-1/2 AFM Heisenberg Hamiltonian on a single hexagonal
plaquette is defined by
\begin{eqnarray}
 H^{\text{\hexagon}}=\!\sum\limits_{n=1}^{6}\left(J_1\, \boldsymbol{S}^n\cdot\boldsymbol{S}^{n+1}
\!+\!J_2\, \boldsymbol{S}^n\cdot\boldsymbol{S}^{n+2}\right),\,
 \boldsymbol{S}^{n+6}\!=\!\boldsymbol{S}^n,
 \label{ham}
\end{eqnarray}
where $\boldsymbol{S}^n$ is the spin-1/2 operator at the $n$-$th$ vertex of the hexagon. 
So, $n$ runs from 1 to 6, in addition to the 
periodic boundary condition (PBC). 
$J_1$ and $J_2$ are the respective NN and NNN exchange interaction strengths. 
$J_1$ and $J_2$ compete against each other 
while computing the minimum value of classical energy of the hexagon, 
which in other words means that $J_2$ invokes frustration in this model 
with respect to $J_1$. Here simultaneous 
minimization of all bond energies fails while constructing the 
classical ground state, which on the other hand generates  
highly degenerate ground state.
A schematic view of this spin model is shown in Fig  \ref{lattice}(a). 
Thermally stable multipartite entanglement is predicted before 
in this model at the extreme frustrated limit, $J_2/J_1$=1 \cite{Moumita1}.

\begin{figure}[h]
\begin{center}
 \includegraphics[width=200pt]{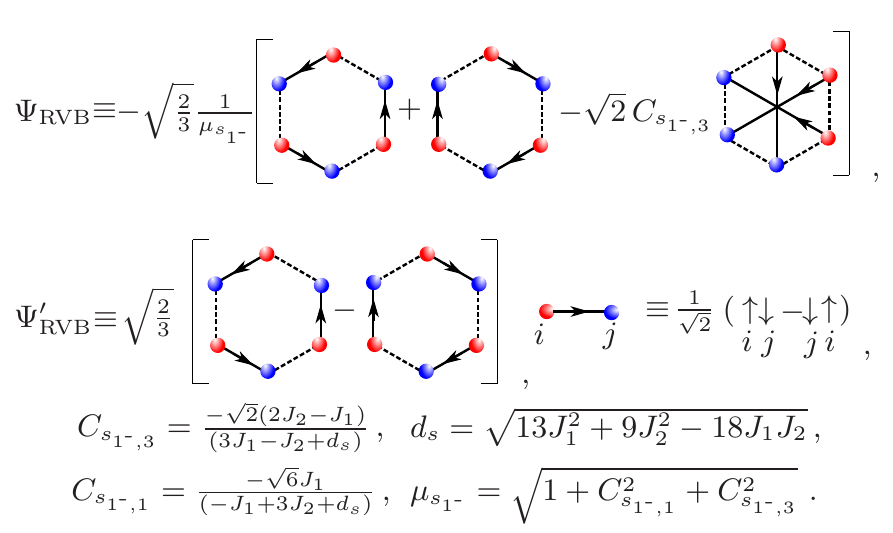}
  \caption{Pictorial representation of $\Psi_{\rm RVB}$ and $\Psi^\prime_{\rm RVB}$. 
Arrow indicates the spin ordering in a singlet dimer bond.}
 \label{RVBs}
\end{center}
\end{figure}

The total spin operator, $\boldsymbol S_{\rm T}=\sum_{n=1}^6\boldsymbol S^n$, 
as well as $z$-component of the total spin, 
$S^z_{\rm T}$, commute with the Hamiltonian, 
$H^{\text{\hexagon}}$, since the system is SU(2) invariant. 
The eigenvalue equation of $H^{\text{\hexagon}}$ 
has been solved exactly by spanning the Hamiltonian matrix 
separately into the subspaces for 
different $S^z_{\rm T}$ values as they are being good quantum numbers. 
The Hilbert space of this six-spin Hamiltonian 
consists of $2^6$ states and those states comprise to
five singlets ($S_{\rm T}=0$), nine triplets ($S_{\rm T}=1$), 
five quintets ($S_{\rm T}=2$) and one septet ($S_{\rm T}=3$). 
The exact analytic expressions of all singlet ($\ket{s}$), 
triplet ($\ket{t}$), quintet ($\ket{q}$) and septet ($\ket{h}$) 
states with energy eigenvalues have been listed in the Appendix \ref{eigensystem}. 
Six pairs of doubly-degenerate states are there. One singlet pair, three triplet pairs 
and two quintet pairs are found to be degenerate.
Among those singlets, only two, $\ket{s_{ 1^\Minus}}$ and $\ket{s_2}$ 
can be expressed in terms of Kekule configurations. 
So, they can be recognized as RVB states for a single hexagon.
Those two particular singlets are denoted 
by $\Psi_{\rm RVB}$ and $\Psi^\prime_{\rm RVB}$, respectively. 
The pictorial views of those states,
$\Psi_{\rm RVB}\;(\ket{s_{ 1^\Minus}})$ and 
$\Psi^\prime_{\rm RVB}\;(\ket{s_2})$ are shown in Fig  \ref{RVBs}.
$\Psi_{\rm RVB}$ is symmetric, whereas, 
$\Psi^\prime_{\rm RVB}$ is antisymmetric under the reflection about the 
mirror planes passing through any 
vertices of the hexagon. Those mirror planes are shown by 
dashed lines noted with $M_1$, $M_2$ and $M_3$ in Fig \ref{lattice} (f). 
While both the RVB states are antisymmetric under spin inversion as well as 
reflection about the mirror planes normal to any 
NN bonds of the hexagon (${\cal M}$-symmetry). 
Those mirror planes are shown by dashed lines marked with 
$M_4$, $M_5$ and $M_6$ in Fig  \ref{lattice} (f). 
Rotation and reflection symmetries of all the eigenstates are 
described in the Appendix \ref{eigensystem}.

\begin{figure}[h]
\begin{center}
  \includegraphics[width=210pt]{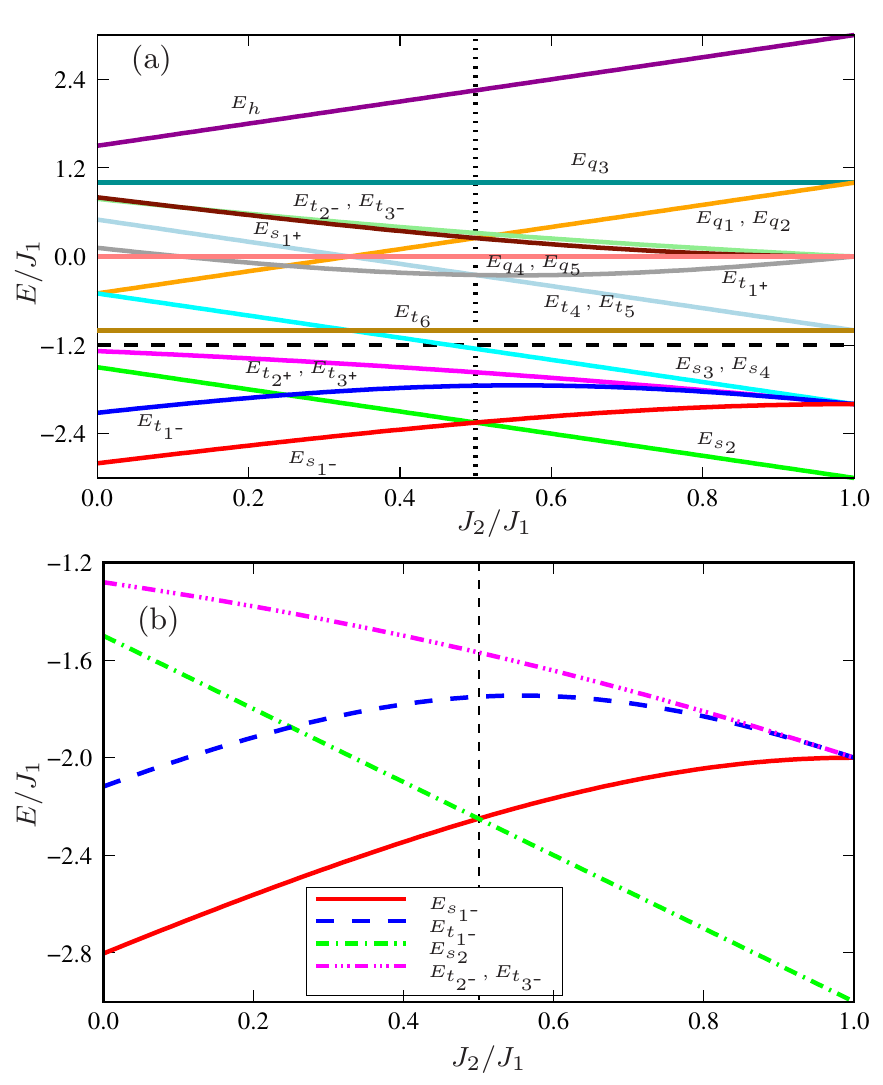}
 \end{center}
   \caption{Variation of energy eigenvalues of single plaquette against $J_2/J_1$.}
     \label{E2D}
  \end{figure}

Variation of all those energy eigenvalues against $J_2/J_1$ is shown in Fig  \ref{E2D} (a). 
Several crossovers among the energy states are found with the change of $J_2/J_1$. 
Energy states below the dashed line are considered to develop the POT. 
An expanded view of the region below the dashed line 
is separately shown in Fig  \ref{E2D} (b).   
The variation of energies for two lowest singlets, $\ket{s_{ 1^\Minus}}$ and $\ket{s_2}$, 
along with that of three lowest triplets, $\ket{t_{ 1^\Minus,\alpha}}$ and doubly degenerate 
$\ket{t_{ 2^\Minus,\alpha}}$, $\ket{t_{ 3^\Minus,\alpha}}$,  
are plotted in Fig  \ref{E2D} (b). Here, $\alpha=x,y,z$, denotes  
the three different components of the triplet state. 
In this region of energy, one singlet-singlet ($\ket{s_{ 1^\Minus}}$-$\ket{s_2}$) 
and one singlet-triplet ($\ket{s_2}$-$\ket{t_{ 1^\Minus,\alpha}}$) 
crossovers are observed.

$ E_{s_{ 1^\Minus}}$ and $ E_{s_2}$ are the energies of the singlet states 
$\ket{s_{ 1^\Minus}}$ and $\ket{s_2}$, respectively. Similarly, 
$ E_{t_{ 1^\Minus}}$ and $ E_{t_{ 2^\Minus}}= E_{t_{ 3^\Minus}}$ are the energies 
of the triplet states $\ket{t_{ 1^\Minus,\alpha}}$ and doubly degenerate 
$\ket{t_{ 2^\Minus,\alpha}}$, $\ket{t_{ 3^\Minus,\alpha}}$, respectively. 
Ground state is always a total spin singlet. 
Two RVB states form the ground states in two different regions, say, R$_1$ and R$_2$. 
In the moderate frustrated region (R$_1$), $0<J_2/J_1<1/2$ where 
$\Psi_{\rm RVB}$ or $\ket{s_{ 1^\Minus}}$ is the ground state, while 
$\Psi^\prime_{\rm RVB}$ or $\ket{s_2}$ is that in the extreme 
frustrated region (R$_2$) when $1/2<J_2/J_1<1$. 
Energies of these two singlets cross themselves at the point $J_2/J_1=1/2$. 
So, ground state is doubly degenerate at this point. 
This figure reveals that two types of spin gaps are there 
for a single hexagonal plaquette, those are associated with the 
transitions between two different sets of lowest energy states. 
One is associated with a singlet-triplet transition (triplet gap)
when $J_2/J_1<1/4$ while other one is a singlet-singlet (singlet gap) 
when $J_2/J_1>1/4$. Three triplet and one singlet states are found 
degenerate at the point $J_2/J_1=1$. 
Based on these lowest energy singlet and triplet states, POT has been 
developed to study the properties of PVBS phase 
of this frustrated honeycomb model.   
In this theoretical development, singlets
$\Psi_{\rm RVB}$ and $\Psi^\prime_{\rm RVB}$ get condensed separately in the 
regions R$_1$ and R$_2$, respectively. 
Thus in the PVBS phase, ground state is actually the 
product of respective PRVB states defined on a regular array of 
plaquettes. One such array is shown in Fig  \ref{lattice} (c). 
As a result, ground state is six-fold degenerate and preserve the 
symmetry of the Hamiltonian in every case \cite{Sachdev1,Misguich}. 
\section{plaquette operator theory}
\label{POT} 
In order to develop the POT, all the six 
spin-1/2 operators within each individual hexagonal plaquette, 
{\em i.e.}, $\boldsymbol{S}^n,\; n=1,2,\cdots , 6$, 
are expressed in terms of the 
creation operators of a number of low energy eigenstates of 
$H^{\text{\hexagon}}$. Those states are chosen from the 
complete list available in the Appendix \ref{eigensystem}. 
As the true Hilbert space of a single plaquette consists of 64 states, 
the corresponding 64 creation operators are defined by the following 
notations. 
 \begin{equation}
 \begin{aligned}
   & \ket{s_j}= s^\dagger_j\ket{0}, \;
    \ket{t_{a,\alpha}}= t^\dagger_{a,\alpha}\ket{0},\;
     \ket{q_{b,\nu}}= q^\dagger_{b,\nu}\ket{0},  \\
    &\ket{q_{b,\alpha}}=q^\dagger_{b,\alpha}\ket{0}, \;
    \ket{h_{\zeta}}= h^\dagger_\zeta\ket{0}, \;
    \ket{h_{\alpha}}= h^\dagger_{\alpha}\ket{0},
     \end{aligned}
 \end{equation}
where $\ket{0}$ denotes the vacuum state. 
All the operators are assumed to satisfy the bosonic commutation relations 
whenever defined for the same plaquette, otherwise commute 
when they are specified for the different plaquettes. 

The alphabets, 
$s,\,t,\,q$ and $h$ stand for the singlet, triplet, quintet and septet states, 
respectively. Subscripts, $j= 1^\pm,2,3,4$, $a=1^\pm,2^\pm, 3^\pm,4,5,6$, and 
$b=1,2,3,4,5$, indicate five singlets, nine triplets and 
five quintets, respectively. Additional subscripts 
$\alpha=x,y,z$, $\nu=1^\pm$, and $\zeta=1^\pm,2^\pm$,  
denote the components of those multiplets. 
The peculier combinations of numbers and signs for $j$ and $a$ are 
found useful to write down the corresponding eigenstates and eigenvalues in 
a compact form. The similar argument does hold in a different way 
for the additional subscripts, $\alpha=x,y,z$, $\nu=1^\pm$, 
and $\zeta=1^\pm,2^\pm$.  
Precisely, the particuler index $\pm$ is used for bunching up a pair of eigenstates 
in a single expression and so the individual sign does not correspond to eigenvalue of parity or any other operators.
Anyway, the completeness relation in this full Hilbert space is thus given by 
\begin{equation}
 \begin{aligned}
 \sum\limits_{j} s^\dagger_j\,s_j  &+ \sum\limits_{a,\alpha} t^\dagger_{a,\alpha}\,t_{a,\alpha} + \sum\limits_{b,\nu} q^\dagger_{b,\nu}\,q_{b,\nu} 
 +\sum\limits_{b,\alpha} q^\dagger_{b,\alpha}\,q_{b,\alpha}
\\&+\sum\limits_{\zeta} h^\dagger_{\zeta}\,h_{\zeta}+
 \sum\limits_{\alpha} h^\dagger_{\alpha}\,h_{\alpha}=1. 
 \label{constraint}
 \end{aligned}
 \end{equation}
The Hamiltonian (Eq \ref{ham}) in the full Hilbert space assumes the form
 \begin{equation}
 \begin{aligned}
H^{\text{\hexagon}}&=\sum_{\substack{j}}E_{s_j}s^\dagger_j\,s_j 
+ \sum_{\substack{a,\alpha}} 
E_{t_a}\,t^\dagger_{a,\alpha}\,t_{a,\alpha}   +\sum\limits_{b,\nu} E_{q_b} \,q^\dagger_{b,\nu}\,q_{b,\nu} \\
& +
 \sum\limits_{b,\alpha} E_{q_b} \,q^\dagger_{b,\alpha}\,q_{b,\alpha}\!
+ E_{h} \sum\limits_{\zeta} h^\dagger_{\zeta}\,h_{\zeta}+
   E_{h} \sum\limits_{\alpha}   h^\dagger_{\alpha}\,h_{\alpha}.
 \end{aligned}
 \end{equation}
However, the spin operators, $S^n_\alpha$, are expressed in the Fock space 
constituted by a limited number of plaquette operators as shown 
below 
 \begin{equation}
 \begin{aligned}
  S^n_\alpha=&A^n_{\eta}\left(t^\dagger_{\eta^\Minus,\alpha}\,s_{ 1^\Minus}+
s^\dagger_{1^\Minus }t_{ \eta^\Minus,\alpha}\right)+
  B^n_{\eta}\left(t^\dagger_{ \eta^\Minus,\alpha}\,s_2+s^\dagger_2\,t_{\eta^\Minus ,\alpha}\right)\\
  &-i\,\epsilon_{\alpha\beta\gamma}\,D^n_{\eta\xi}\,t^\dagger_{ \eta^\Minus,\beta}\,t_{ \xi^\Minus,\gamma}.
  \label{ope}
 \end{aligned}
 \end{equation}
Here, $n$, again denotes the position of spin within a plaquette, 
$\alpha=x,y,z$, $\eta=1,2,3$, and $\xi=1,2,3$. 
The matrix elements, 
$A^n_{\eta}=\langle s_{1^\Minus}|S^n_\alpha|t_{\eta^\Minus ,\alpha}\rangle$, 
$B^n_{\eta}=\langle s_{2}|S^n_\alpha|t_{\eta^\Minus ,\alpha}\rangle$, and 
$D^n_{\eta\xi}=\langle t_{\xi^\Minus ,\gamma} |S^n_\alpha|t_{\eta^\Minus ,\beta}\rangle$,  
are given in the Appendix \ref{lambda}. The reduced space is spanned 
by the two lowest singlets, ($\ket{s_{ 1^\Minus}}$,$\ket{s_2}$),  
and three lowest triplets, ($\ket{t_{ 1^\Minus,\alpha}}$, $\ket{t_{ 2^\Minus,\alpha}}$, 
$\ket{t_{ 3^\Minus,\alpha}}$), those are shown in Fig  \ref{E2D} (b). 
Therefore, the spin commutation relations, 
$[S^n_\alpha,S^m_\beta]=i\epsilon_{\alpha\beta\gamma}\,\delta_{nm}\,S^n_\gamma$,  
will be preserved by taking into account the completeness relation in the 
truncated Hilbert space, which reads as,
 \begin{eqnarray}
 \sum\limits_{j=1^\Minus,2} s^\dagger_j\,s_j  + 
\sum\limits_{\eta,\alpha} t^\dagger_{\eta^\Minus,\alpha}\,t_{\eta^\Minus,\alpha}=1. 
 \label{constraint2}
 \end{eqnarray}
The form of $S^n_\alpha$ that is given in Eq \ref{ope} has been used to 
express the inter-plaquette interactions in the AFM Heisenberg Hamiltonian 
for the honeycomb lattice. Thus, the low-energy dynamics of bosonic 
version of this model 
will be studied in this truncated Hilbert space. 
Nevertheless, a more general form of the spin operators, $S^n_\alpha$ 
in terms of all singlets and triplets are available in the 
 Appendix \ref{lambda}.
 \section{The Low energy BOSON MODEL}
 \label{LEBM}
In this section, POT has been employed to study the PVBS phase of 
the $J_1$-$J_2$ AFM Heisenberg honeycomb model. 
The non-Bravais honeycomb lattice is assumed as a 
triangular lattice composed of hexagonal plaquettes as shown 
in Fig \ref{lattice}(b).  
The Hamiltonian is expressed in terms of spin operators 
those are assigned to a definite site of a particular plaquette  
which is constituted by six different sites. 
\begin{equation}
 \begin{aligned}
 H\!=&\sum\limits_{i}\!\big[H^{\text{\hexagon}}_{\boldsymbol{r}_i}\!
+\!J_1\!\left(\boldsymbol{S}^1_{\boldsymbol{r}_i}\!\cdot\!\boldsymbol{S}^4_{\boldsymbol{r}_i+\boldsymbol{\tau}_1}\!
+\!\boldsymbol{S}^2_{\boldsymbol{r}_i}\!\cdot\!\boldsymbol{S}^5_{\boldsymbol{r}_i+\boldsymbol{\tau}_1+\boldsymbol{\tau}_2}\!+\!
 \boldsymbol{S}^3_{\boldsymbol{r}_i}\!\cdot\!\boldsymbol{S}^6_{\boldsymbol{r}_i+\boldsymbol{\tau}_2}\right)\\
 &+J_2\,\big(\boldsymbol{S}^1_{\boldsymbol{r}_i}\!\cdot\boldsymbol{S}^5_{\boldsymbol{r}_i+\boldsymbol{\tau}_1}\!+\!\boldsymbol{S}^1_{\boldsymbol{r}_i}\!\cdot\boldsymbol{S}^3_{\boldsymbol{r}_i+\boldsymbol{\tau}_1} 
+ \boldsymbol{S}^1_{\boldsymbol{r}_i}\!\cdot\boldsymbol{S}^5_{\boldsymbol{r}_i+\boldsymbol{\tau}_1+\boldsymbol{\tau}_2}\\
&+\boldsymbol{S}^2_{\boldsymbol{r}_i}\!\cdot\boldsymbol{S}^4_{\boldsymbol{r}_i+\boldsymbol{\tau}_1}
 +\boldsymbol{S}^2_{\boldsymbol{r}_i}\!\cdot\boldsymbol{S}^6_{\boldsymbol{r}_i+\boldsymbol{\tau}_1+\boldsymbol{\tau}_2}
+\boldsymbol{S}^2_{\boldsymbol{r}_i}\!\cdot\boldsymbol{S}^4_{\boldsymbol{r}_i+\boldsymbol{\tau}_1+\boldsymbol{\tau}_2}\\
&+ \boldsymbol{S}^2_{\boldsymbol{r}_i}\!\cdot\boldsymbol{S}^6_{\boldsymbol{r}_i+\boldsymbol{\tau}_2}
+\boldsymbol{S}^3_{\boldsymbol{r}_i}\!\cdot\boldsymbol{S}^5_{\boldsymbol{r}_i +\boldsymbol{\tau}_1+\boldsymbol{\tau}_2}
 +\boldsymbol{S}^3_{\boldsymbol{r}_i}\!\cdot\boldsymbol{S}^1_{\boldsymbol{r}_i+\boldsymbol{\tau}_2}\\
&+\boldsymbol{S}^3_{\boldsymbol{r}_i}\!\cdot\boldsymbol{S}^5_{\boldsymbol{r}_i+\boldsymbol{\tau}_2}
 +\boldsymbol{S}^4_{\boldsymbol{r}_i}\!\cdot\boldsymbol{S}^6_{\boldsymbol{r}_i+\boldsymbol{\tau}_2}
+\boldsymbol{S}^6_{\boldsymbol{r}_i}\!\cdot\boldsymbol{S}^4_{\boldsymbol{r}_i+\boldsymbol{\tau}_1}\big)\big].
 \label{ham1}
 \end{aligned}
  \end{equation}
  Here, the vector $\boldsymbol{r}_i$ indicates the position of a particular plaquette 
while the other two vectors, $\boldsymbol{\tau}_1$ and $\boldsymbol{\tau}_2$ 
are used to point the positions of surrounding plaquettes in the resulting 
triangular lattice. $\boldsymbol{S}^n_{\boldsymbol{r}}$ denotes the spin-1/2 operator 
at the $n$-$th$ vertex of the hexagonal plaquette at the position ${\boldsymbol{r}}$. 
Therefore, 
in this case, $\boldsymbol{\tau}_1$ and $\boldsymbol{\tau}_2$ could be considered as 
the primitive vectors in this effective triangular lattice 
formed by the hexagonal plaquettes. They can be expressed in the following way: 
 \begin{equation}
 \boldsymbol{\tau}_1=3\,a\,\hat{y} \,\quad \textrm{ and} \,
\quad \boldsymbol{\tau}_2=\frac{3\sqrt{3}\,a}{2} \, 
\hat{x}-\frac{3\,a}{2} \, \hat{y},
 \end{equation}
where $a$ is the NN lattice spacing of the original honeycomb 
lattice which is henceforth assumed to be unity.
The Hamiltonian $H$ is SU(2) invariant. Ultimately, $H$ is expressed in terms of 
singlet and triplet plaquette operators when the $\boldsymbol{S}^n_{\boldsymbol{r}}$ 
is replaced by bosonic plaquette operators using the Eq \ref{ope}, 
and thus has the following form:  
   \begin{equation}
  \begin{aligned}
  H=E_0+H_{02}+H_{20}+H_{30}+H_{21}+H_{40}+H_{22}.
     \end{aligned}
     \label{ham2}
 \end{equation}
 Obviously, $H_{nm}$ indicates different terms in the Hamiltonian in which 
$E_0$ is a constant. The expression of relevant terms in the momentum space 
will be shown in the next section. In $H_{nm}$, $n$ and $m$ indicate the numbers of 
triplet and singlet operators, respectively. 
Expressions of $H_{nm}$ with non-zero value of $m$ will be different for the regions R$_1$ and R$_2$ those are 
introduced before in the Sec \ref{SHP}. Value of $E_0$ will be different in the regions R$_1$ and R$_2$. 
It should be noted that the Hamiltonian is expressed in terms of the two lowest energy singlets and 
three lowest energy triplets only. The contribution of 
higher energy singlets and triplets as well as all quintets and the septet is neglected. 
So, the truncated form of the relevant constraint (Eq \ref{constraint2}) 
has been taken into account by adding
the following term to the Hamiltonian (Eq \ref{ham2}), 
   \begin{equation}
  \begin{aligned}
 -\mu\sum\limits_{i}\bigg( \sum\limits_{j=1^\Minus,2} s^\dagger_{j,i}\,s_{j,i}  + 
\sum\limits_{\eta,\alpha} t^\dagger_{\eta^\Minus,i,\alpha}\,t_{\eta^\Minus,i,\alpha}-1\bigg), 
     \end{aligned}
 \end{equation}
where $\mu$ is the Lagrange multiplier. The summation index $i$ runs over the all triangular lattice sites.
 Here $\mu$ can be imagined as the chemical potential which is assumed to be site independent in accordance 
with the translational invariance of the system. 

To study the low-energy dynamics of this system, the effective Hamiltonian is derived by 
condensing the lowest energy RVB states, $\ket{s_{1^-,i}}$ and $\ket{s_{2,i}}$ in every site $i$ 
for the respective parameter regimes R$_1$ and R$_2$, separately. 
Thus, for the implementation of plaquette operator formalism,  
one of the two singlet states $\ket{s_{j,i}},\;j=1^-,2$, is assumed to be condensed
and so has been substituted by a number, $\bar s$ in Eq \ref{ham2}. 
The effect of condensation is thus taken into account by making the following replacement, 
$s^\dagger_{j,i}=s_{j,i}=\langle s^\dagger_{j,i}\rangle=\langle s_{j,i}\rangle=\bar s$ \cite{Sachdev2}. 

As a result, the effective Hamiltonian contains the operators related to 
the singlet $\ket{s_{2,i}}$, 
($\ket{s_{1^-,i}}$) in R$_1$, (R$_2$) along with the three triplets. 
Now the value of the constant, $E_0$ is given by the equation, 
$E_0=N^\prime\left[\bar s^2E-\mu\left(\bar s^2-1\right)\right]$, in which 
$E=E_{s_{ 1^\Minus}}$ ($E_{s_{2}}$) for the region R$_1$ 
(R$_2$). $N^\prime=N/6$ where $N$ is the total 
number of sites of the original honeycomb lattice. 
Fourier transformation of the operators $t^\dagger_{\eta,i,\alpha}$ and $s^\dagger_{j,i}$ are 
     \begin{equation}
  \begin{aligned}
t^\dagger_{\eta,i,\alpha}&=\frac{1}{\sqrt{N^\prime}}\sum\limits_{\bold{k}}\text{exp}\left(-i\bold{k}\cdot\boldsymbol{R}_i\right)t^\dagger_{\eta,\bold{k},\alpha},\\
s^\dagger_{j,i}&=\frac{1}{\sqrt{N^\prime}}\sum\limits_{\bold{k}}\text{exp}\left(-i\bold{k}\cdot\boldsymbol{R}_i\right)s^\dagger_{j,\bold{k}}. 
     \end{aligned}
 \end{equation}
  Here, the momentum sum runs over the BZ of the triangular lattice.
\section{Mean-field Analysis}
\label{MFA}
In order to estimate the ground state energy, 
$E_{\rm G}$ and the singlet to triplet 
spin gap, $\Delta$, of the honeycomb model, a mean-field theory has been developed. 
Triplet dispersion relations based on the PVBS ground state are obtained. 
Hamiltonian retains upto the quadratic terms, 
as a result, the terms $H_{30}, H_{21}, H_{40},H_{22}$ 
have been neglected. However, expressions of 
those terms in the momentum space are available in the Appendix \ref{mean}. 
Now the mean-field Hamiltonian becomes,   
 \begin{equation}
  \begin{aligned}
   H_{\rm MF} =  E_0+H_{02}+H_{20}.
   \label{ham3}
  \end{aligned}
\end{equation}
This truncated Hamiltonian is capable to capture the 
low energy dynamics of the system valid at the low temperatures. 
The expressions of $H_{02}$ and $H_{20}$ in terms of singlet and triplet 
operators in momentum space become
    \begin{equation}
  \begin{aligned}
  H_{02}=\sum\limits_{\bold{k}}\left(E_{s_m}-\mu\right)s^\dagger_{m,\bold{k}}s_{m,\bold{k}}, 
     \end{aligned}
 \end{equation}
  with $m=2$ and $1^\Minus$ for the regions R$_1$ and R$_2$, respectively.  
   \begin{equation}
  \begin{aligned}
  H_{20}\!=\!\!\!\sum\limits_{\bold{k},\eta,\xi}\!\!\!\! \!X^{\eta\xi}_{\bold{k}} \,t^\dagger_{\eta^\Minus,\bold{k},\alpha}t_{\xi^\Minus,\bold{k},\alpha}
  \!+\!\!\!\frac{Y^{\eta\xi}_{\bold{k}}}{2}\!\!\left(\!t^\dagger_{\eta^\Minus,\bold{k},\alpha}t^\dagger_{\xi^\Minus,-\bold{k},\alpha}\!\!+\!
  t_{\eta^\Minus,-\bold{k},\alpha}t_{\xi^\Minus,\bold{k},\alpha}\!\right), 
     \end{aligned}
 \end{equation}
  where, $\eta,\xi= 1,2,3$ and $\alpha = x,y,z$. The coefficients $X^{\eta\xi}_{\bold{k}}$ 
and $Y^{\eta\xi}_{\bold{k}}$ are given in Appendix \ref{mean}. 
$H_{02}$ is already diagonalized in the singlet basis space where the 
singlet excitation energy is $\Omega_s=\left( E_{s_m}-\mu\right)$.
The six-component vector $\Psi^\dagger_{\bold{k},\alpha}
=\left (t^\dagger_{1^\Minus,\bold{k},\alpha} t^\dagger_{2^\Minus,\bold{k},\alpha} t^\dagger_{3^\Minus,\bold{k},\alpha}
t_{1^\Minus,\bold{- k},\alpha} t_{2^\Minus,\bold{- k},\alpha} t_{3^\Minus,\bold{- k},\alpha}\right)$
 is introduced to diagonalize $H_{20}$ in the basis space 
comprised of three different triplets.
Thus, the matrix form of Eq \ref{ham3} looks like 
 \begin{equation}
  \begin{aligned}
   H_{\rm MF}=E^\prime_0+H_{02}+\frac{1}{2}\sum\limits_{\bold{k}}\Psi^\dagger_{\bold{k},\alpha}{H_\bold{k}}\Psi_{\bold{k},\alpha},
   \label{ham4}
  \end{aligned}
 \end{equation}
where 
 \begin{equation}
 \begin{aligned}
E^\prime_0=E_0-\frac{3}{2} \sum\limits_{\bold{k}} \sum_{\substack{\eta= 1,\\ 2 , 3 }}   X^{\eta\eta}_{\bold{k}},
 \end{aligned}
 \end{equation}
and $H_\bold{k}$ is a $6\times 6$ matrix. $H_\bold{k}$ has the following 
form when expressed in terms of two symmetric submatrices, $X_\bold{k}$ 
and $Y_\bold{k}$. 
\begin{equation}
 \begin{aligned}
H_\bold{k}=
 \left( 
 { \begin{array}{cc}
  X_\bold{k} & Y_\bold{k} \\
  Y_\bold{k} & X_\bold{k} \\
  \end{array}}
\right).
 \end{aligned}
 \end{equation}
The elements of $X_\bold{k}$ 
and $Y_\bold{k}$ are $X^{\eta\xi}_\bold{k}$ and $Y^{\eta\xi}_\bold{k}$, respectively, 
where all of them are real. In addition, all the respective 
off-diagonal elements of $X_\bold{k}$ and $Y_\bold{k}$ are 
the same. After diagonalization, the Hamiltonian assumes the form \cite{Colpa}
\begin{equation}
  \begin{aligned}
   H_{\rm MF}=E_{\rm G}+H_{02}+\frac{1}{2}\sum\limits_{\bold{k}}\Phi^\dagger_{\bold{k},\alpha}{H^\prime_\bold{k}}\Phi_{\bold{k},\alpha}, 
  \end{aligned}
 \label{ham5}
\end{equation}
where the expression of ground state energy is 
\begin{equation}
  \begin{aligned}
   E_{\rm G}=E_0+\frac{3}{2}\sum\limits_{\eta,\bold{k}}\left(\Omega_{\eta,\bold{k}}-X^{\eta\eta}_{\bold{k}}\right). 
  \end{aligned}
\end{equation}
The diagonalized matrix in this case looks like 

$H^\prime_\bold{k}=
 \left( 
 { \begin{array}{cc}
  h_\bold{k} &  0 \\
  0 &  -h_\bold{k} \\
  \end{array}}
\right)$  
where
$  h_\bold{k} = \left( 
 { \begin{array}{ccc}
  \Omega_{1,\bold{k}} &  0 & 0 \\
  0 & \Omega_{2,\bold{k}} &0 \\
  0 & 0 & \Omega_{3,\bold{k}} 
  \end{array}}
\right).$ 
Again, each triplet dispersion, $\Omega_{\eta,\bold{k}},\,\eta=1,2,3$, 
is triply degenerate, since the Hamiltonian (Eq \ref{ham1}) is SU(2) invariant. 
The eigenvectors $\Phi^\dagger_{\bold{k},\alpha}$ is given by 
$\Phi^\dagger_{\bold{k},\alpha}=\left (b^\dagger_{1^\Minus,\bold{k},\alpha} b^\dagger_{2^\Minus,\bold{k},\alpha} b^\dagger_{3^\Minus,\bold{k},\alpha}
b_{1^\Minus,\bold{- k},\alpha} b_{2^\Minus,\bold{- k},\alpha} b_{3^\Minus,\bold{- k},\alpha}\right)$.
Two sets of boson operators $t_{\eta^-}$ and $b_{\eta^-}$, ($\eta=1,2,3$),  
are connected to each other by the following relation \cite{Colpa},  
\begin{equation}
  \begin{aligned}
   \Phi_{\bold{k},\alpha}= M_{\bold{k}}   \Psi_{\bold{k},\alpha},\quad \rm{ where} \quad
    M_\bold{k}=
 \left( 
 { \begin{array}{cc}
  U^\dagger_\bold{k} & -V^\dagger_\bold{k} \\
  -V^\dagger_\bold{k} & U^\dagger_\bold{k} \\
  \end{array}}
\right).
  \end{aligned}
\end{equation}
Coefficients of the  $3 \times 3$ Hermitian matrices 
$U^\dagger_\bold{k}$ and $ V^\dagger_\bold{k}$ are the 
Bogoliubov coefficients $u^{\eta\xi}_\bold{k}$ and $v^{\eta\xi}_\bold{k}$, respectively.
The analytic expressions of the triplet excitation energies 
$\Omega_{\eta,\bold{k}}$ and the Bogoliubov
coefficients $u^{\eta\xi}_\bold{k}$ and $v^{\eta\xi}_\bold{k}$ written in terms of 
the components $X^{\eta\xi}_\bold{k}$ and $Y^{\eta\xi}_\bold{k}$ are 
available in Appendix \ref{mean}.

Two self-consistent equations for the determination of 
mean-field parameters, $\mu$ and $\bar s^2$ are obtained by minimizing the 
ground state energy, $E_{\rm G}$, with respect to themselves as 
$\frac{\partial E_{\rm{G}}}{\partial \mu}=0$, and 
$\frac{\partial  E_{\rm{G}}}{\partial \bar s^2}=0$. The resulting equations are 
\begin{equation}
  \begin{aligned}
&\mu=E+\frac{3}{2 N^\prime}\sum\limits_{\eta,\bold{k}}
\left[\frac{\partial \Omega_{\eta,\bold{k}}}{\partial \bar s^2}-
\frac{Y^{\eta\eta}_{\bold{k}}}{\bar s^2}\right], \\
&\bar s^2 =1+\frac{3}{2 N^\prime}\sum\limits_{\eta,\bold{k}}
\left[\frac{\partial \Omega_{\eta,\bold{k}}}{\partial \mu}+1\right]. 
  \end{aligned}
\end{equation}
Again, $m=2$ and $1^\Minus$ for the regions R$_1$ and R$_2$, respectively.  
By substituting the numerical values of $\mu$ and $\bar s^2$ 
the singlet, $\Omega_{s_m}= E_{s_m}-\mu $, and the three triplet,
 $\Omega_{\eta,\bold{k}},\,\eta=1,2,3$, excitation energies have been obtained. 
Value of $\bar s^2$ is always positive as expected and less than unity, while 
 $\mu$ is always negative. The singlet excitations, $\Omega_{s_{m}}$ are found to be 
always dispersionless in the mean-field approximation. 
The self-consistent equations do not converge in the region 
$0.0\leq J_2/J_1\leq 0.11$. This type of non-convergence 
in the mean-field procedure for the plaquette and bond-operator theories  
have been reported before \cite{Doretto,Susobhan}. 
Triplet dispersion along with the evolution of topological nodes are 
discussed in the next section (Sec \ref{TN}). 

\begin{figure}[h]
\begin{center}
   \includegraphics[width=230 pt]{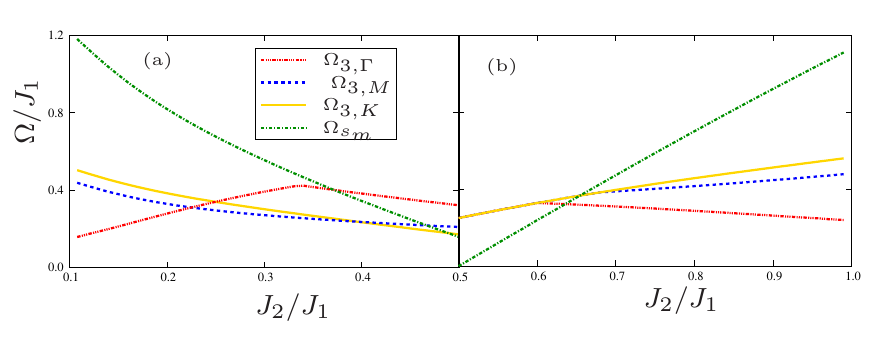}
 \end{center}
   \caption{ Variation of excitation energies for 
 $ \Omega_{3,\Gamma}$, $ \Omega_{3,M}$ and $ \Omega_{3,K}$ 
with respect to $E_{\rm{G}}$ against $J_2/J_1$ (a) in region R$_1$,
   (b) in region R$_2$.}
     \label{gap}
  \end{figure}

The value of $\Delta$ has been obtained by measuring the energy difference 
between the ground and the lowest triplet states. 
Energy of the triplet dispersion, $\Omega_{3,\bold{k}}$, is always lower than 
those of other triplets, $\Omega_{1,\bold{k}}$ and $\Omega_{2,\bold{k}}$. 
Variation of $\Omega_{3,\bold{k}}$ in the BZ indicates that 
minima of $\Omega_{3,\bold{k}}$ happens to occur at  
the symmetric points $\Gamma$, M and K in the BZ. 
Thus variation of excitation energies for 
$ \Omega_{3,\Gamma}$, $ \Omega_{3,M}$ and 
$ \Omega_{3,K}$ with respect to $J_2/J_1$ have been plotted 
in in Figs \ref{gap} (a) and \ref{gap} (b) for the regions R$_1$ and R$_2$, 
respectively, along with that of $ \Omega_{s_m}$. 
Those energies are measured with respect to the 
ground state energy. 
By comparing the energies of $ \Omega_{3,\Gamma}$, $ \Omega_{3,M}$ and 
$ \Omega_{3,K}$ in the region R$_1$, 
it is evident that $\Omega_{3,\Gamma}$,  $ \Omega_{3,M}$ 
and $ \Omega_{3,K}$ are the lowest 
when $0.11<J_2/J_1<0.23$, $0.23<J_2/J_1<0.40$ and $0.40<J_2/J_1<0.50$, 
respectively. 
In region R$_2$, $ \Omega_{3,\Gamma}$, $ \Omega_{3,M}$ and 
$ \Omega_{3,K}$ have the same value for $0.5<J_2/J_1< 0.66$, 
and thereafter $ \Omega_{3,\Gamma}$ is the lowest. 
The value of $\Delta$ has been estimated form this comparative study. 

 \begin{figure}[h]
   \centering
    \includegraphics[width=230pt]{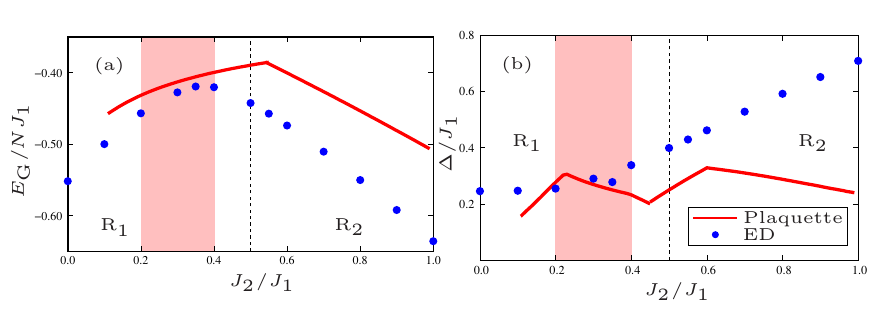}
   \caption{Variation of (a) Ground-state energy per site and (b) spin gap against $J_2/J_1$. The shaded region indicates the PVBS phase.}
     \label{Epla}
  \end{figure}    

 Variation of ground state energy per site with respect to
$J_2/J_1$ is shown in Fig  \ref{Epla}(a). Result based on the POT is plotted in 
red line and that has been compared with the exact diagonalization 
data for $N$=32 sites shown in blue points \cite{Sindzingre}. The
value of $E_{\rm{G}}/NJ_1$ shows closer agreement with the 
exact diagonalization data in the disordered PVBS  
phases which is identified by the region shaded in pink, 
where this formalism stands valid.
However, the ground state energy is always higher than 
the true value because of the fact
that POT is basically a variational approach. 
$E_{\rm{G}}/NJ_1$ shows significant departure from the
exact diagonalization data in the ordered regions since 
POT fails to capture the quantum correlation in those regions.
Similarly, spin gap ($\Delta/J_1$) has been evaluated through the 
POT and that is shown in Fig  \ref{Epla}(b). 
The estimated value of spin gap in the PVBS phase (shaded in pink) 
is significantly close to the numerical values. In addition
to that, gap is also found in the magnetic ordered phases 
which are supposed to be 
gapless. $\Delta/J_1$ corresponds to the gap
between the ground state energy and the minimum of the triplet excitations. 
More accurate estimation of the ground state energy and the spin gap 
can be made in this formalism by accommodating the 
higher energy modes of the single plaquette excitations in the POT those 
are neglected before. 
 \section{Triplet dispersions and topological modes} 
\label{TN}
\begin{figure*}
\includegraphics[width=505pt]{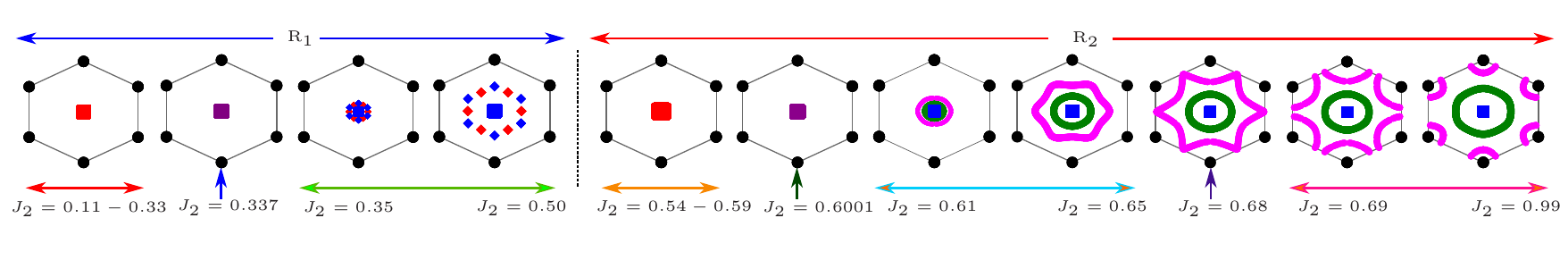}
\caption{Kaleidoscope of the imprints of evolutionary nodes and nodal lines on the BZ 
with the change of $J_2/J_1$, where $J_1=1$.}
\label{node}
\end{figure*} 

Two different kinds of band-touching points or nodes, 
are noted depending on the number of meeting bands. 
They are termed as two-band and three-band touching points (TBTP), 
where two and three bands are found to meet there, respectively. 
Two types of two-band touching points are 
identified depending on their nature of dispersion 
relation around the respective touching points. 
Those nodes are called Dirac and QBTP. 
For the QBTP, energy of triplet excitation is proportional to 
square of the momentum in the vicinity of the touching point. 
The QBTP can be regarded as a pair of Dirac nodes \cite{Soljacic}. 
Similarly, two types of TBTP are 
identified for the same reason as stated before. 
It has been noted that energy of triplet excitation of all the 
three meeting bands are proportional to the 
square of the momentum in the neighborhood of the touching point 
for the case of TBTP in the region R$_1$. 
On the other hand, the lowest energy band is flat 
for the TBTP found in the region R$_2$. 
Energy of triplet dispersion of the remaining 
two meeting bands are proportional to the 
square of the momentum near the touching point in this case.  
Emergence and evolution of those point nodes as well 
as the DLNs  
with the variation of $J_2/J_1$ have been shown 
in Fig \ref{node}. Dirac, QBTP and TBTP appear both in the 
regions R$_1$ and R$_2$, while DLN and flat band appear only 
in the region R$_2$. 
\begin{figure}[h]
\centering
\includegraphics[width=250pt]{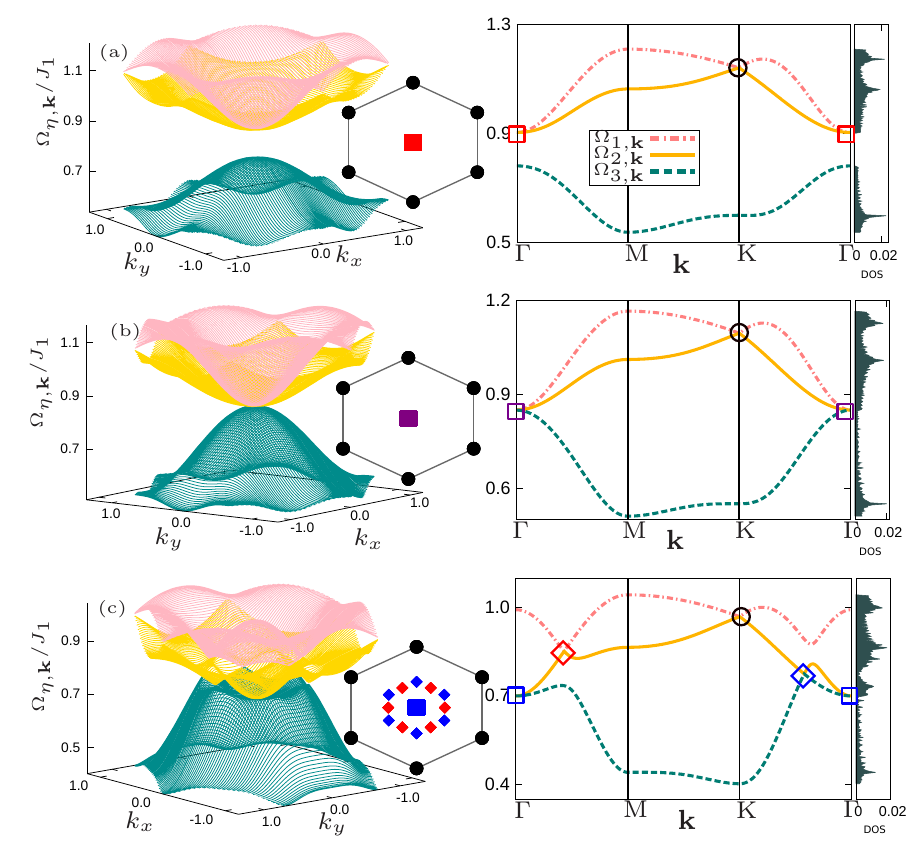}
 \caption{Evolution of nodes in region R$_1$ with $J_2/J_1$,
 $J_2/J_1=$ (a) 0.30, (b) 0.337, (c) 0.45. }
\label{triplet1}
\end{figure} 
 
The 3D plot of the triplet dispersions 
for the region R$_1$ have been shown in 
Fig \ref{triplet1}(a)-(c). Those figures are 
supplemented by the respective dispersion along the high-symmetry
pathway $(\Gamma$,M,K,$\Gamma)$, density of states (DOS), as well as the 
location of topological nodes within the BZ. 
DOS is useful to estimate the values of band gap and band width.   
The number of band touching points changes with $J_2/J_1$, 
however, a Dirac node is always formed 
due to the band touching of upper two bands at the K point 
regardless the values of $J_2/J_1$, which is denoted by black circle 
in the figures. So, this particular node is 
protected by the symmetry of the Hamiltonian, while other 
nodes appear as a result of accidental degeneracy. 

\begin{figure}[h]
 \includegraphics[width=230pt]{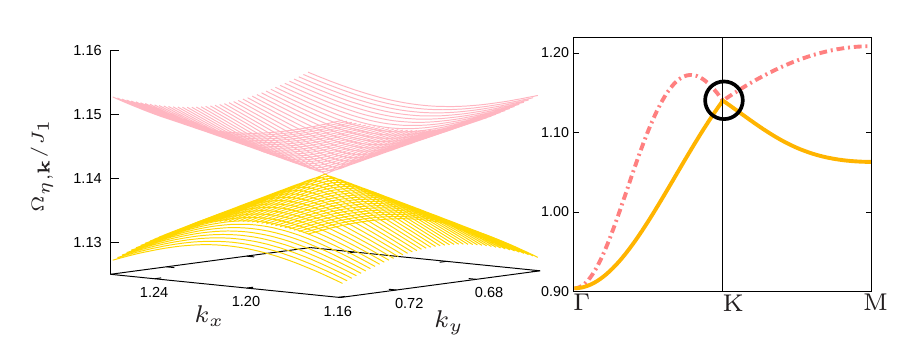}
\caption{Closer view of triplet excitations around the 
Dirac node when $J_2/J_1=0.30$.}
     \label{dirac_node}
  \end{figure}

Closer view of this Dirac node is shown in Fig 
\ref{dirac_node}. 
This particular Dirac node 
is analogous to that appeared in graphene \cite{Wallace}. 
Thus it can be regarded as a generic feature of the 
honeycomb lattice. 
The same Dirac node is found in the triplet magnon excitation 
of the FM Heisenberg model in the collinear phase, though it is 
absent in the AFM case \cite{Boyko}. 
This discrepancy is attributed to the fact that 
FM ground state does not break the ${\cal M}$-symmetry, 
while the AFM ground state does \cite{Boyko}. 
It is worth mentioning in this situation that 
FM state is the exact ground state of the Hamiltonian, while the 
AFM state is not the exact one. 
On the other hand, under the same mirror reflection, both  
$\Psi_{\rm RVB}$ and $\Psi^\prime_{\rm RVB}$ states are antisymmetric. 
Further, the PVBS ground states preserve the 
symmetry of the Hamiltonian. 
As a result, this particular Dirac node in the triplet dispersion is present 
in both the regions R$_1$ and R$_2$, irrespective of the 
values of exchange strengths. 

  \begin{figure}[h]
 \includegraphics[width=230pt]{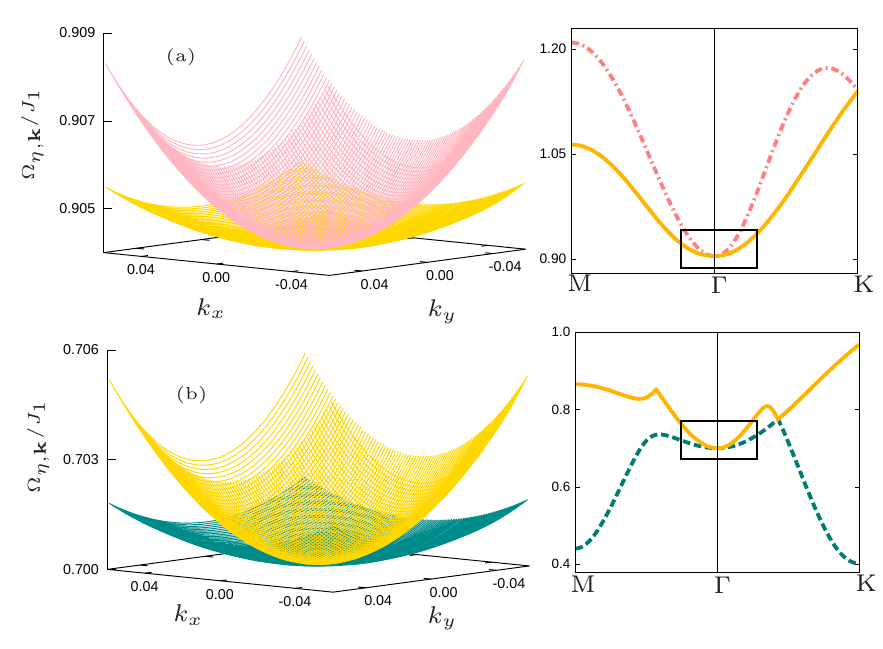}
\caption{Closer view of triplet excitations around the QBTPs 
(a) $J_2/J_1=0.30$ and (b) $J_2/J_1=0.45$.}
     \label{QBTP}
  \end{figure} 

\begin{figure}[h]
 \includegraphics[width=250pt]{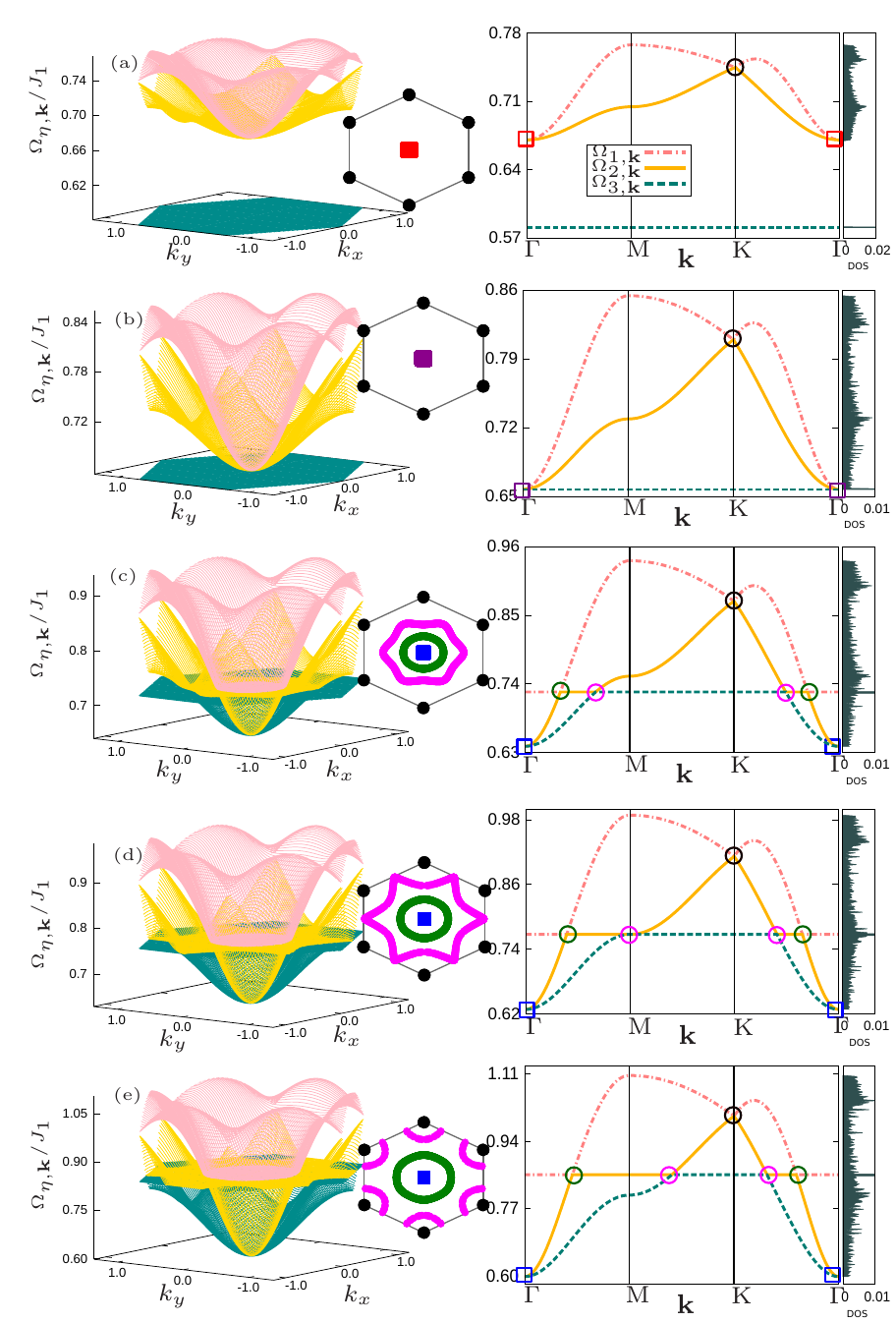}
\caption{Evolution of nodes and nodal lines in region R$_2$ with $J_2/J_1$,
 $J_2/J_1=$ (a) 0.55, (b) 0.6007, (c) 0.65, (d) 0.68, (e) 0.75. }
     \label{triplet2}
  \end{figure} 

The TBTP is noticed at the $\Gamma$ point 
only when $J_2/J_1=0.337$, whereas a QBTP  
between upper two band is found at that point when $J_2/J_1<0.337$, 
by leaving a gap between the lower two bands. 
This QBTP always appears at the $\Gamma$ point irrespective 
of the values of $J_2/J_1$, as long as $J_2/J_1<0.337$. 
On the other hand, six pairs of Dirac nodes are found 
to appear with equal share between the lower and upper two bands,  
as soon as $J_2/J_1>0.337$, along with the emergence of 
another QBTP between the lower two bands at the $\Gamma$ point. 
These additional Dirac nodes are found to emerge 
in the immediate vicinity of the $\Gamma$ point, 
while the QBTP originates at the $\Gamma$ point itself. 
Although the Dirac nodes shift towards the M (upper Dirac nodes) 
and K (lower Dirac nodes)
points with the increase of $J_2/J_1$, the QBTP does not change its position. 
This picture is valid for the region, $0.337<J_2/J_1<0.50$.  
Locations of those movable nodes in the BZ and dispersion along 
$(\Gamma$,M,K,$\Gamma)$ pathway are marked by red and 
blue diamonds, for the Dirac nodes in between upper and lower 
two bands, respectively. 
Therefore, the TBTP (purple square) at 
the $\Gamma$ point for $J_2/J_1=0.337$ is replaced by 
QBTP between upper bands (red square) and that between lower 
bands (blue square) for 
$J_2/J_1<0.337$ and $J_2/J_1>0.337$, respectively.  
Closer view of triplet excitations around those QBTPs are shown 
in Fig \ref{QBTP} (a) and (b), at two definite values, 
$J_2/J_1=0.30$ and $J_2/J_1=0.45$, respectively. 
No band gap is found in the regime, $0.337\leq J_2/J_1<0.50$. 

    \begin{figure}[h]
 \includegraphics[width=230pt]{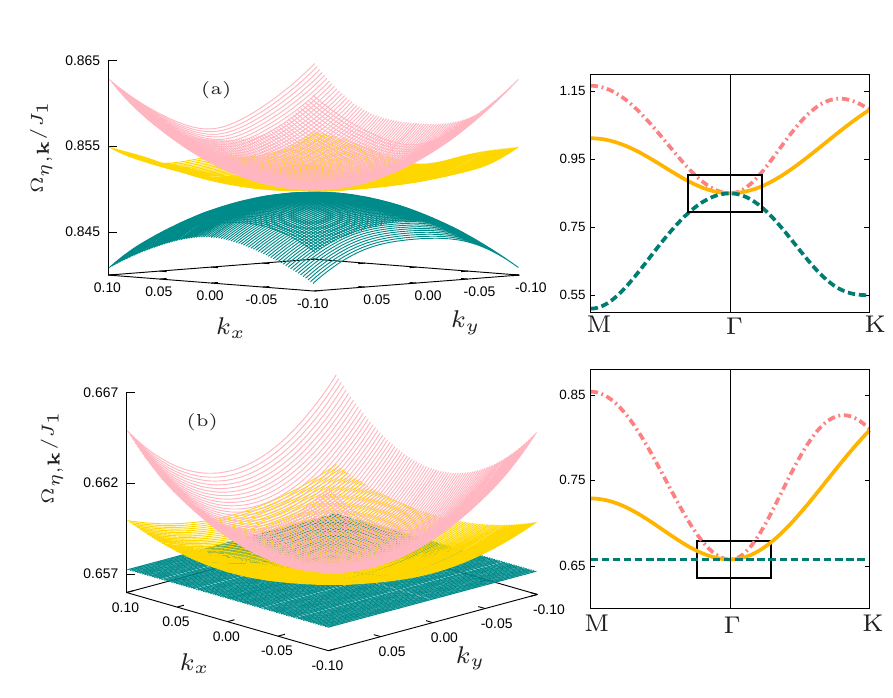}
\caption{Closer view of triplet excitations around the TBTPs  
for (a) $J_2/J_1=0.337$ and (b) $J_2/J_1=0.6001$. }
     \label{triple_point}
  \end{figure}  

As a result, the fixed Dirac node at the K point (black circle)
always appears at every vertex of the hexagonal BZ with 
the coordinates,  
 $(0,\frac{4\pi}{9})$, $(0,\Minus\frac{4\pi}{9})$, 
$(\frac{2\pi}{3\sqrt{3}},\frac{2\pi}{9})$, 
$(\Minus\frac{2\pi}{3\sqrt{3}},\frac{2\pi}{9})$, 
$(\frac{2\pi}{3\sqrt{3}},\Minus\frac{2\pi}{9})$ 
and $(\Minus\frac{2\pi}{3\sqrt{3}},\Minus\frac{2\pi}{9})$,  
for any values of $J_2/J_1$.
So, the QBTP (red square) always appears at the center 
of that ($\Gamma$ point) when $J_2/J_1<0.337$. 
This  situation is as shown in Fig  \ref{triplet1}(a) for $J_2/J_1=0.30$. 
When $J_2/J_1=0.337$, 
QBTP between upper bands (red square) at the center of BZ is replaced by 
the triple point (purple square), 
which is shown in Fig \ref{triplet1}(b). 
Closer view of the energy dispersion in the vicinity of 
TBTP is shown in Fig \ref{triple_point} (a). 
Finally, the TBTP (purple square) is replaced by 
another QBTP between lower bands (blue square) 
at the center of BZ when $J_2/J_1>0.337$. 
The movable Dirac nodes appear symmetrically around the center 
of BZ as shown in Fig \ref{triplet1}(c) for $J_2/J_1=0.45$. 

Similarly in the region R$_2$, 3D plots of the triplet excitations, $\Omega_{\eta,\bold{k}}/J_1$,
covering the BZ, as well as along the one-dimensional 
pathway, have been shown in 
Fig \ref{triplet2} (a)-(e), for five  
different values of $J_2/J_1=$ (a) 0.55, (b) 0.6007, (c) 0.65, (d) 0.68, 
(e) 0.75. 
The fixed Dirac node located at the K point is present as usual like before 
for any value of $J_2/J_1$.
The dispersion relation depicted 
in Fig \ref{triplet2} (a) is similar to that shown in 
Fig \ref{triplet1} (a), with the exception that the lowest band is 
flat in the region R$_2$. The band gap is larger in this case. 
The emergence of TBTP point 
is noted like before, and this time it occurs for the value $J_2/J_1=0.6001$,  
which is shown by the purple square. Again it appears at the $\Gamma$ point. 
A magnified view close to this touching point is shown 
in Fig \ref{triple_point} (b). 
In region R$_2$, the lowest excitation 
till remains dispersionless, however, up to $J_2/J_1=0.61$.
Thus, the dispersion relation presented  
in Fig \ref{triplet2} (b) can be compared to 
that in Fig \ref{triplet1} (b) in the same fashion. 

The nature of dispersion relation for the region R$_2$ changes dramatically 
beyond the TBTP with the increase of $J_2/J_1$. 
In this case, two DLNs appear where one between upper and another between 
lower two bands instead of the six pairs of Dirac nodes as found before in the region R$_1$.  
Both the nodal lines are closed and appear at the same energy. 
Among the two associated bands, one is always flat for each of those two DLNs, 
which means that magnitude of DOS at this value of energy 
is extremely high. DLNs shift towards higher energies 
with the increase of $J_2/J_1$. 

\begin{figure}[h]
 \includegraphics[width=230pt]{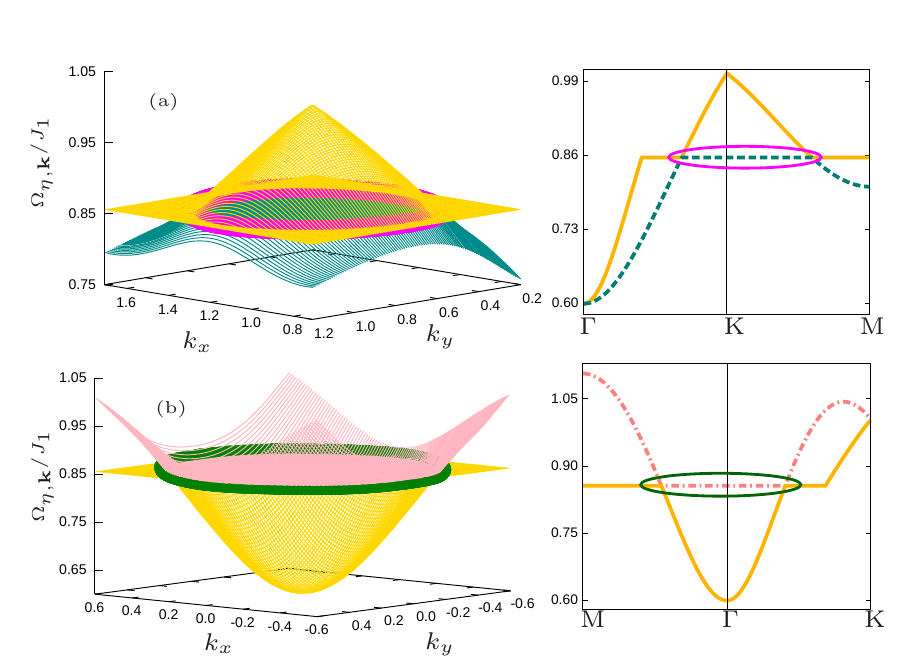}
\caption{Closer view of triplet excitations around the two circular 
nodal lines for $J_2/J_1=0.75$.}
     \label{nodal_line}
  \end{figure}

Nodal line formed between the upper two bands is circular with 
the $\Gamma$ point at its center, while that 
between the lower two bands is hexagonal and symmetric around 
the $\Gamma$ point. 
In the beginning, both the DLNs are found in the 
immediate vicinity of $\Gamma$ point. 
With the increase of 
$J_2/J_1$, both the nodal lines move away from the $\Gamma$ point 
but with different fashions. 
The radius of the circular nodal line increases, 
while the hexagonal nodal line changes 
its shape and becomes K-centered circular one when $J_2/J_1$ crosses the value 0.68. 
The structural deformation in the later case takes place when $J_2/J_1=0.68$, or 
as soon as this nodal line touches the M point otherwise. 
Radius of this second circular nodal line decreases with the further increase of 
$J_2/J_1$ beyond 0.68.  
The evolution of two DLNs can be found in the Figs \ref{triplet2} (c) - (e). 
Fig \ref{nodal_line} shows the magnified view of the two circular nodal lines 
centered around the K and $\Gamma$ points, when $J_2/J_1=0.75$. 
Nonetheless, both the DLNs are always symmetric around the center of BZ ($\Gamma$ point), 
which corresponds to the fact that they are 
topologically protected by the ${\cal PT}$-symmetry 
invariance of the system. 
Which means that these DLNs are of type II \cite{Fang,Mertig}. 
However, all through the region, $0.61\leq J_2/J_1\leq 0.99$, 
a QBTP node between lower two bands (blue square) is found to present 
at the $\Gamma$ point, which is again similar to the previous case. 
Fig \ref{node} contains the kaleidoscope of distinct patterns of topological nodes 
formed within the BZ, over the whole parameter region, $0.11\leq J_2/J_1\leq 0.99$. 
\section{Topological Phases}
\label{TP}
The system studied in terms of triplet operators is topologically trivial, 
since the Hamiltonian does not break the ${\cal T}$-symmetry.   
However, in this section, emergence of nontrivial topology 
in the R$_1$ region will be discussed in the presence of 
${\mathcal H}_P$, which is SU(2) invariant and 
breaks the ${\cal T}$-symmetry.  
It also satisfies an additional criterion, which states that 
the respective off-diagonal elements in the 
submatrices $X_\bold{k}$ and $Y_\bold{k}$ 
may be the same, barring at least one. 
This additional criterion is obtained empirically  
and found necessary for the nontriviality in this case. 
\begin{equation}
 \begin{aligned}
{\cal H}_P&=\sum\limits_{\bold{k}}i D_1\left(t^\dagger_{2^\Minus,\bold{k},\alpha}t_{3^\Minus,\bold{k},\alpha} -t^\dagger_{3^\Minus,\bold{k},\alpha}t_{2^\Minus,\bold{k},\alpha}\right)\\ \nonumber
&+i\frac{D_2}{2}\left(t^\dagger_{2^\Minus,\bold{k},\alpha}t^\dagger_{3^\Minus,-\bold{k},\alpha} -t_{2^\Minus,-\bold{k},\alpha}t_{3^\Minus,\bold{k},\alpha}\right). 
 \end{aligned}
\label{hhexa}
 \end{equation}
The total Hamiltonian including ${\cal H}_P$ is SU(2) invariant,  
but breaks the ${\cal T}$-symmetry in this three-band system, since 
$H_\bold{k} \neq H_{-\bold{k}}^{\ast}$ \cite{Arghya}. 
The presence of ${\cal H}_P$ lifts the degeneracy at the 
band touching points, at the same time, Berry curvature is deformed  
in such a way that non-zero Chern number emerges when $D_1 \neq D_2$. 
The condition, $D_1 \neq D_2$ implies one dissimilar term 
among the respective off-diagonal elements in the 
submatrices $X_\bold{k}$ and $Y_\bold{k}$, 
which ultimately satisfies the additional criterion in other words.  
Different topological phases appear with 
the variation of $D$'s in the region R$_1$. 

In order to draw a topological phase diagram, 
values of C for distinct triplet energy bands are obtained for 
every topological phase. The value of C of a particular band has been calculated by
integrating the Berry curvature over the BZ, 
\begin{equation}
  \begin{aligned}
   {\rm C}=\frac{1}{2\pi }\iint_{\rm BZ} F(\bold{k})dk_x dk_y,
  \end{aligned}
\label{Chern}
\end{equation}
where the Berry curvature of that band, $F(\bold{k})$, is expressed 
as, 
$A_{\mu}(\bold{k}) = \braket{n(\bold{k})|\partial_{k_{\mu}}|n(\bold{k})}$ as 
$F (\bold{k})= {\partial_{ k_x}} A_{y}(\bold{k})-
{\partial_{ k_y}} A_{x}(\bold{k})$, and 
$|n(\bold{k})\rangle$ is the eigenvector of that particular triplet band 
in the $S^z_{\rm T}=0$ sector. However, the result will 
remain unchanged if the eigenvectors with $S^z_{\rm T}=\pm$ sectors 
are taken into account instead, since the total Hamiltonian is still  
SU(2) invariant. 
Value of C is obtained by evaluating the integral, 
Eq \ref{Chern} numerically \cite{Fukui}. 
The number of edge states is related 
with the non-zero Chern numbers which confirms the existence of 
nontrivial topological phase. This relation is governs by the BEC  
rule. 

To calculate the edge state spectrum PBC imposed along the $\hat{y}$ 
direction is removed. Which leaves a strip  
of honeycomb lattice having $N$ plaquettes 
along the $\hat{y}$ direction. 
The system is still assumed infinitely long along the $\hat{x}$ direction. 
A replica of this structure is shown in Fig \ref{lattice} (f).
By applying the Fourier transformation on the bosonic operators 
only along the $\hat{x}$ direction, the Hamiltonian matrix of 
order $2N\times2N$ for the resulting system has been obtained. 
Energies of triplet exciations for bulk-edge states are obtained 
by diagonalizing the Hamiltonian matrix numerically. 

\begin{figure*}
    \centering
   \includegraphics[width=500pt]{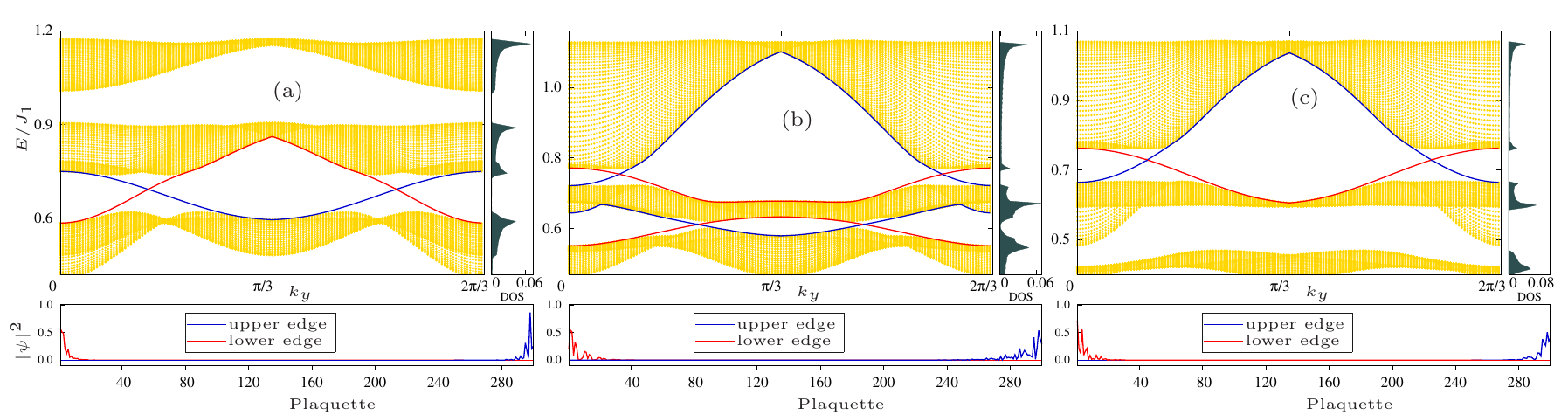}
  \caption{Edge states of the plaquette honeycomb lattice. The side panel shows the DOS.
 The lower panel indicates variation of probability density of both edge
modes with plaquette number at a fixed value of $k_y$.}
 \label{edge}
   \end{figure*}

Bulk-edge dispersion relations for three distinct topological phases of the system 
have been shown in Fig \ref{edge} in the one-dimensional BZ. 
Energy dispersion obtained for $J_2/J_1=0.30$, $D_1/J_1=0.15$ 
and $D_2/J_1=0.5$ is shown in Fig \ref{edge} (a). 
The state of the system corresponds to topological 
phase with C=$(2,\Minus 2,0)$.
Topological phases with C=$(2,\Minus 4,2)$ and C=$(0,\Minus 2,2)$ 
appear when $D_2/J_1=\Minus 0.7$ and $\Minus 0.8$, respectively. 
Energy dispersion of those two topological phases 
are shown in Fig \ref{edge} (b) and (c), respectively. 
In every case, edge-state modes are found to appear 
in accordance to the BEC rule \cite{Hatsugai}. 

\begin{figure*}
\includegraphics[width=508pt]{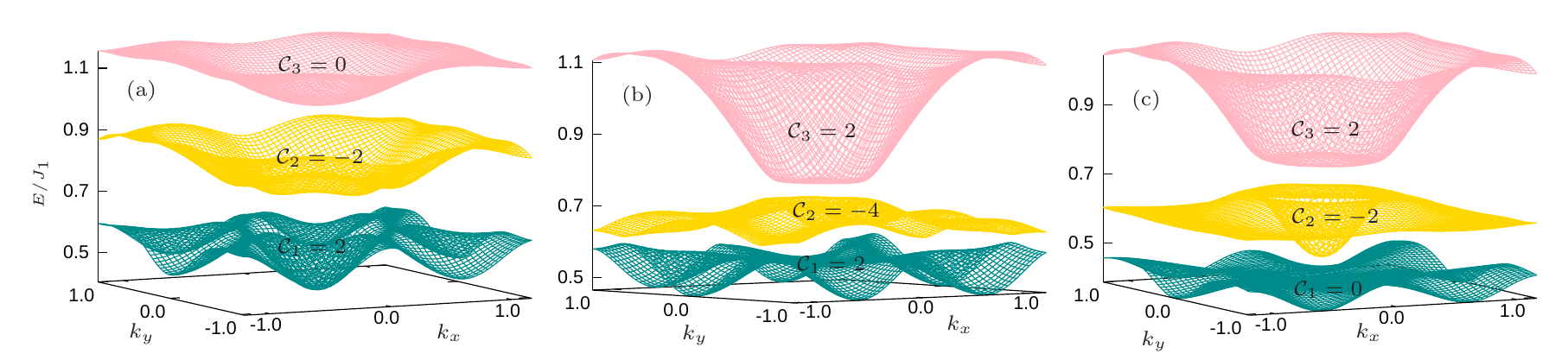}
\caption{The triplet dispersion relations after adding the ${\cal H}_P$ term. 
Chern numbers of the respective bands are stated for three different topological phases.}
  \end{figure*}

\begin{figure}[h]
 \includegraphics[width=230.0 pt]{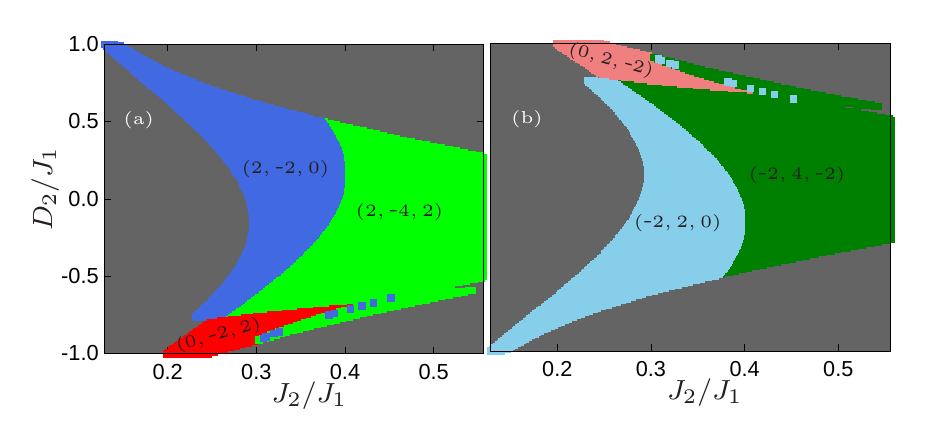}
  \caption{Topological phase diagram for (a) ${D_1}/{J_1}=0.15$, (b) ${D_1}/{J_1}=-0.15$.}
     \label{chern}
  \end{figure}

Total six distinct topological phases have been found in this system. 
The remaining three topological phases can be obtained in the 
following way. For a fixed $J_2/J_1$, if a particular topological 
phase with C=$(n_1,n_2,n_3)$ appears at definite values of $D$'s, then, 
another topological phase with C=$(\Minus n_1,\Minus n_2,\Minus n_3)$ 
must appear upon reversal of signs of $D$'s but keeping 
their values fixed. Hence Chern numbers are 
found to reverse their signs with the reversal of signs of $D$'s.
This phenomenon is depicted in the topological 
phase diagrams of the system as shown in Figs \ref{chern} (a) and (b),  
where the values of ${D_1}/{J_1}$ are kept fixed at $\pm0.15$, respectively.  

Moreover, other topological phases apart from those six may 
appear with different choices of ${\cal T}$-symmetry breaking terms. 
But an arbitrary choice of ${\cal H}_P$ may ultimately 
results in nonphysical complex eigenenergies for the following reasons. 
Eigenenergies are obtained via the bosonoic 
Bogoliubov transformation where the 
product of $I_B$ and $H_{\bold{k}}$ is being diagonalized instead of 
$H_{\bold{k}}$ alone, (Appendix \ref{mean}). This product is always  
a non-Hermitian matrix when the Hermitian ${\cal H}_P$ with complex elements 
is added to the real $H_{\bold{k}}$. With some exception, non-Hermitian matrix  
generally leads to complex eigenvalues. 
In this study, values of $D_1$ and $D_2$ in ${\cal H}_P$ are
chosen in such a way that real eigenenergies are obtained. 
Surprisingly, no topological phase 
appears in the region R$_2$ by any choice of ${\cal H}_P$. 

Most of the two-band systems like fermionic Haldane and Kitaev models 
formulated on the 
honeycomb lattice exhibit an unique topological phase, 
C=$\pm 1$, as well as the bosonic FM Heisenberg models including 
the NNN DMI term and the combination of NN Kitaev and SAI terms
reveals the same phase in the presence of external magnetic field 
\cite{Haldane,Kitaev,Owerre1,Joshi}. 
Existence of that particular phase has been verified experimentally 
in the three different cases among four of them \cite{Jotzu,Yokoi,Chen}. 
In contrast, this three-band system exhibits multiple 
topological phases. In comparison to other  
magnetic systems, topological phases emerge 
in this case in the triplet excitations with respect to 
a spin-disordered ground state where the system is SU(2) invariant. 
Further, coexistence of spin gap and topological phases is found in this model, 
while for the other bosonic systems topological phases are obtained in the 
absence of spin gap.  However, in the Kitaev model, topological phase 
emerges on the spin-liquid ground state in the presence of spin gap 
when the magnetic field is non-zero \cite{Kitaev}. 

\begin{figure}[h]
  \includegraphics[width=230pt]{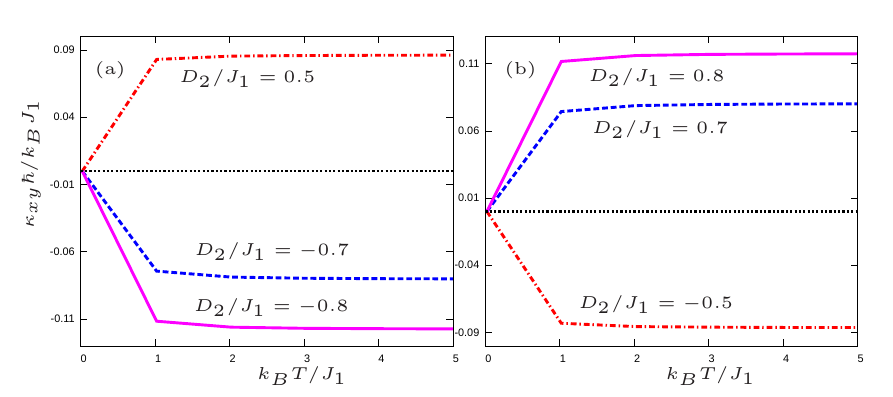}
\caption{Variation of $\kappa_{xy}$ with $T$ for (a) $D_1/J_1=0.15$, (b) $D_1/J_1=-0.15$, when $J_2/J_1=0.30$.}
     \label{Hall1}
  \end{figure}

In order to study the topological phase transition,  
value of thermal Hall conductance, $\kappa_{xy}$, has been computed. 
$\kappa_{xy}$ of the system can be expressed in 
terms of $F(\bold{k})$ as \cite{Murakami}, 
   \begin{equation}
  \begin{aligned}
   \kappa_{xy}(T)=-\frac{k^2_B T}{4\pi^2\hbar}\,\sum\limits_{n}\, \iint_{BZ} 
c(\rho_{n}(\bold{k}))\,F_{n}(\bold{k})\,dk_x dk_y, 
  \end{aligned}
\end{equation}
where $n$ is the band index. 
$T$ is the temperature, $k_B$ is the Boltzmann constant 
and $\hbar$ is the reduced Planck's constant. 
$F_n(\bold{k})$ is the Berry curvature of the $n$-th band. 
$c(x)=(1+x)\left(\ln{\frac{1+x}{x}}\right)^2-\left(\ln x\right)^2 -2{\rm Li}_2(-x)$, where
${\rm Li}_2(z)=-\int_{0}^{z} du \frac{\ln{(1-u)}}{u}$, 
and $ \rho_{n}(\bold{k})$ is the Bose-Einstein distribution function, {\em i.e.},
$\rho_{n}(\bold{k})=1/(e^{E^n_{\bold{k}}/k_{B}T} -1)$. 
Value of $\kappa_{xy}(T)$ does not change at high temperature region 
and it is different for different topological phase. 
The variation of $\kappa_{xy}(T)$ with $T$ for six different 
topological phases is shown in Fig \ref{Hall1}.
  
\begin{figure}[h]  
     \centering
 \includegraphics[width=230.0 pt]{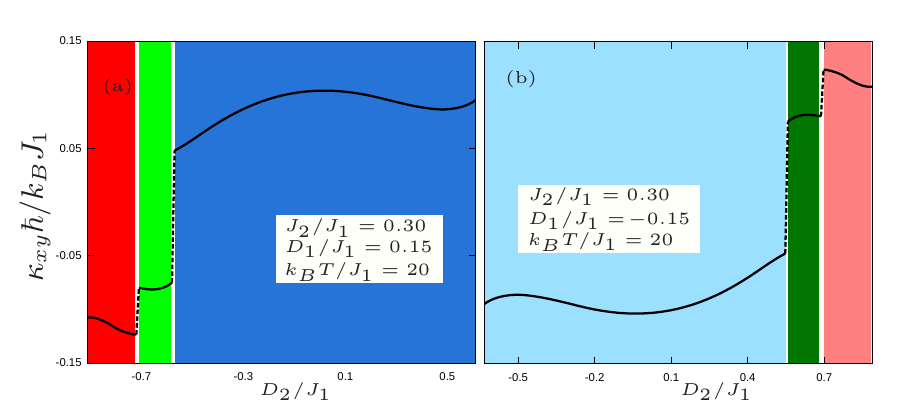}
  \caption{Variation of $\kappa_{xy}$ with $D_2/J_1$ when $k_B T/J_1=20$, 
$J_2/J_1=0.30$ and $D_1=\pm 0.15$ in (a) and (b), respectively. 
Phase transition occurs at the points 
where $\kappa_{xy}$ suffers a sudden jump.}
   \label{Hall2}
  \end{figure}
  
The variation of $\kappa_{xy}$ in the parameter space has been studied 
for fixed value of $T$ to identify the transition among 
the various topological phases. When the system crosses topological 
phase boundary there is a discontinuity in $\kappa_{xy}$. 
Fig \ref{Hall2} (a) and (b) depict topological phase transition 
of the system with respect to $D_2/J_1$ for $D_1/J_1=\pm 0.15$ 
at $J_2/J_1=0.30$.
\section{discussion}
\label{discussion}
In this investigation, emergence of a variety of multiple topological 
nodes is noted as well as a pair of DLNs and a flat band in the triplet dispersions 
of the three-band $J_1$-$J_2$ AFM Heisenberg model on the honeycomb lattice on the basis of 
spin-disordered ground state throughout the whole frustrated 
parameter regime, $0<J_2/J_1<1$. The spin-disordered state is known as PVBS phase 
which actually prevails in the region $0.2<J_2/J_1<0.4$,  
as a true ground state of the system. 
In order to estimate the ground state energy and spin gap, POT has been 
developed on the basis of a limited number of low energy 
exact eigenstates of a single 
$J_1$-$J_2$ AFM Heisenberg hexagon, where all the exact eigenstates have been 
derived indeed. Expressions for the eigenvalues and eigenstates 
are given in the Appendix \ref{eigensystem}. 
Two different PRVB states turn out as the ground states of a single  
hexagon plaquette in the two separate regions R$_1$ 
and R$_2$, while it is doubly degenerate at the meeting point 
of the two regions.  
Thus, POT has been developed on the two regions separately 
based on the respective ground states. The values of 
ground state energy and spin gap obtained via POT is very close 
to the numerical estimations where PVBS phase persists. 
But beyond the region $0.2<J_2/J_1<0.4$, POT overestimates 
the ground state energy. More accurate estimation is 
possible if higher energy eigenstates are taken into account 
in the POT, however, validity of POT is questionable in the 
spin-ordered regions as well. 
It must be noted at this point that although the 
ground state properties of this frustrated model has been 
studied before by using several methods \cite{Albuquerque,Baskaran,
Meng,Sondhi,Jafari,Lamas,Bishop,Ganesh,Zhu,Gong,Oitmaa,Fouet}, 
but an extensive investigation on the topological properties 
is not yet undertaken. 

Topological nodes in the forms of Dirac point, QBTP, 
and two different types of TBTPs are found to appear in this single model 
including a pair of degenerate DNLs in the triplet dispersion bands 
with respect to the spin-disordered ground state. 
Those nodes appear in this three-band system with the variation of $J_2/J_1$. 
Thus, evolution of those point and line nodes 
in this system can be regarded as 
a result of variation of frustration within the system. 
One particular Dirac node among 
all of them is found to bear the reminiscence of honeycomb lattice, 
since its feature is similar to that 
appears before in graphene and Heisenberg honeycomb model 
in the collinear FM phase. Other nodes are tunable. Weyl nodes are found 
in the collinear FM phase of the Heisenberg honeycomb model 
when the NNNN interaction with strength above the 
critical value is taken into account \cite{Boyko}. 
On the other hand, the lack of invariance in 
the collinear ground state under the 
${\cal M}$-symmetry operation bars the emergence of Dirac node 
in the AFM case. 
All the point and line nodes in the triplet dispersions 
emerge with respect to the PVBS ground state with spin 
gap and when the system does not break the 
${\cal PT}$-symmetry. 

Additionally, the system hosts six distinct topological phases when 
a specific ${\cal T}$-symmetry breaking term is included. 
Incorporation of the symmetry breaking terms 
like, Zeeman, DMI, Kitaev and SAI, those are found to exist  
within the materials fail to induce 
nontriviality in this system. 
The reason behind this failure attributes to the fact that here 
POT is formulated in a direction where the system does not break the 
SU(2) invariance in the every intermediate stage. 
Whereas, those symmetry breaking terms do not preserve 
SU(2) but retain the U(1) symmetry. Therefore, 
a modified version of POT on the basis of spin-singlet 
ground state requires  
which is valid for those systems 
where SU(2) is lost but U(1) symmetry is preserved. 
In the modified version of the POT, triple degeneracy of 
the every triplet dispersion will be broken 
leading to 18$\times$18 Hamiltonian matrix, $H_\bold{k}$ 
in Eq \ref{ham4}, with the inclusion of other symmetry broken terms. 
So, the corresponding submatrices $X_\bold{k}$ and $Y_\bold{k}$ are 
of dimension 9$\times$9. In this situation, $H_\bold{k}$ 
must break the ${\cal T}$-symmetry but emergence of 
nontrivial topology may demand additional criteria. 
\section{ACKNOWLEDGMENTS}
We are grateful to Prof. P. Sindzingre for providing us the 
numerical data of ground
state energy. MD acknowledges the UGC fellowship, No. 524067 (2014), India. 
AKG acknowledges a BRNS-sanctioned research project, 
No. 37(3)/14/16/2015, India.


 \appendix
\begin{widetext}
\section{ENERGY EIGENVALUES AND EIGENSTATES OF THE HEISENBERG HEXAGON}
\label{eigensystem}
In this appendix, expressions of all the eigenvectors ($|\nu\rangle$) and 
eigenvalues ($E_\nu$) of the Heisenberg  Hamiltonian for 
a single hexagonal plaquette (Eq \ref{ham}) are given by solving the 
eigenvalue equation, $H^{\text{\hexagon}}|\nu\rangle=E_\nu|\nu\rangle$. 
$H^{\text{\hexagon}}$ possesses the symmetry of a regular hexagon, 
which is studied in terms of a group of 12 elements, known as 
dihedral group $D_6$. $D_6$ is composed of six rotations, $\hat{R}_n$ 
and six reflections, $\hat{M}_n$, $n=1,2,3,4,5,6$. 
$\hat{R}_n$ be the successive $\hat{R}$ operation by 
$n$ times, where $\hat{R}$ implies 
the rotation by $\pi/3$ about the center of the hexagon, 
as depicted in Fig  \ref{lattice} (e). Six different mirror planes for 
 $\hat{M}_n$ operations are shown by dashed lines in Fig  \ref{lattice} (f). 

For the counter clockwise rotation by $\pi/3$, 
the rotational operator, $\hat{R}$, can be defined as  
$\hat{R}\ket {S_1S_2S_3S_4S_5S_6}=\ket{S_2S_3S_4S_5S_6S_1}$, where 
$\ket{S_1S_2S_3S_4S_5S_6}=\ket{S_1^z}\otimes\ket{S_2^z}\otimes\ket{S_3^z}
\otimes\ket{S_4^z}\otimes\ket{S_5^z}\otimes\ket{S_6^z}$, in which 
$\ket{S_n^z}$ is the spin state at the $n$-$th$ vertex.
Obviously $\hat{R}_6$ is the identity 
operation which leaves any state 
unaltered. Each eigenstate of the Hamiltonian, $|\nu\rangle$
has a definite rotational property, 
which can be described in terms of an eigenvalue equation, like 
$\hat{R}_p|\nu\rangle=\lambda_r|\nu\rangle$, where  
$\lambda_r$ be the eigenvalue of the rotational operator $\hat{R}_p$. 
The value of $p$ corresponds to the minimum number 
$\hat{R}$ operations on a definite state unless 
$\lambda_r$ assumes the value either $+1$ or $-1$. 
Obviously, for the same state $\lambda_r$ is always $+1$ for $2p$ number of 
 $\hat{R}$ operations. 
The states with $\lambda_r=+1$ have even parity (symmetric) 
while those with $\lambda_r=-1$ have odd parity (antisymmetric). 
It is found that, every eigenstate has definite values 
of both $p$ and $\lambda_r$, and subsequently has definite parity. 
36 states have even parity while the remaining 28 states have odd parity. 
Values of $p$ and $\lambda_r$ for all eigenstates are shown   
in the Table I. It is observed that 
$p$ takes up  either 1 or 3 and never 
takes up 2, 4 and 5.  For $\Psi_{\rm RVB}$, $\lambda_r=-1$ and $p=1$, while,  
for $\Psi_{\rm RVB}'$, $\lambda_r=1$ and $p=1$. Thus, 
$\Psi_{\rm RVB}'$ does not change sign 
under any number of $\hat{R}$ operations, while $\Psi_{\rm RVB}$ changes sign 
for odd numbers of $\hat{R}$ operations. So,
$\Psi_{\rm RVB}$ is antisymmetric, whereas, 
$\Psi^\prime_{\rm RVB}$ is symmetric under the rotation by the angle $\pi/3$. 

Similarly, the effect of reflections of the eigenstates can be studied in 
terms of an eigenvalue equation 
$\hat{ M}_n|\nu\rangle=\lambda_{{M}_n}|\nu\rangle$. Obviously, $\hat{M}_n^2$ 
is the identity operation which on the otherhand fixes the 
values of $\lambda_{{M}_n}$ to be $\pm 1$ in this case. 
The operations $\hat{ M}_n$ are defined as
\begin{equation}
 \begin{aligned}
\hat{M}_1\ket {S_1S_2S_3S_4S_5S_6} &=\ket{S_1S_6S_5S_4S_3S_2},\quad
\hat{M}_2\ket {S_1S_2S_3S_4S_5S_6} =\ket{S_3S_2S_1S_6S_5S_4},\\
\hat{M}_3\ket {S_1S_2S_3S_4S_5S_6}&=\ket{S_5S_4S_3S_2S_1S_6},\quad
\hat{M}_4\ket {S_1S_2S_3S_4S_5S_6}=\ket{S_6S_5S_4S_3S_2S_1},\\
\hat{M}_5\ket {S_1S_2S_3S_4S_5S_6}&=\ket{S_2S_1S_6S_5S_4S_3},\quad
\hat{M}_6\ket {S_1S_2S_3S_4S_5S_6}=\ket{S_4S_3S_2S_1S_6S_5}. 
\end{aligned}
\end{equation}

\begin{table}[h]
\begin{center}
\label{table_eigenvalues}
\caption{Energy and other eigenvalues of the eigenstates of spin-1/2  Heisenberg hexagon}
\def\arraystretch{0.78}
  \tabcolsep4pt\begin{tabular}{|c|c|c|c|c|c|c|c|c|c|c|}
\hline
 
 $\boldsymbol{S_{\rm T}}$ & $\boldsymbol{S^z_{\rm T}}$ &\textbf{Energy eigenvalues}& $\boldsymbol{\lambda_r}$ & $\boldsymbol{p}$ & $\boldsymbol{\lambda_{ M_1}}$ & $\boldsymbol{\lambda_{ M_2}}$ & $\boldsymbol{\lambda_{ M_3}}$ & $\boldsymbol{\lambda_{ M_4}}$ & $\boldsymbol{\lambda_{ M_5}}$ & $\boldsymbol{\lambda_{ M_6}}$\\  
\hline
 0 & 0 & $ E_{s_{1^\pm}}= -J_1\pm \frac{1}{2}d_s $ & -1 & 1 &1 &1 &1 &-1&-1&-1 \\
\hline
0 & 0 & $ E_{s_2}=-\frac{3}{2}\left(J_1+J_2\right) $  & 1 & 1 &-1 &-1 &-1 &-1&-1&-1 \\
\hline
0 & 0 & $ E_{s_3}= -\frac{1}{2}\left(J_1+3J_2\right) $  & -1 & 3 &1 & & & & &-1  \\
\hline
0 & 0 &$ E_{s_4} = -\frac{1}{2}\left(J_1+3J_2\right) $  & -1 & 3 &-1 & & & & &1  \\
\hline
1 &  &$E_{t_{ 1^\pm,\alpha}}=-J_1\pm \frac{1}{2}d_{t_1}$ & 1 & 1&1 &1 &1 &1&1&1 \\
\hline
1 & 0 &$E_{t_{ 1^\pm,z}}=-J_1\pm \frac{1}{2}d_{t_1}$ & 1 & 1&1 &1 &1 &1&1&1 \\
\hline
 1 & &$E_{t_{ 2^\pm,\alpha}}  =  -\frac{1}{4}(J_1+3J_2\mp d_{t_2})$& 1 & 3&1 & & & & &1\\
\hline
1 & 0 &$E_{t_{ 2^\pm,z}}  =  -\frac{1}{4}(J_1+3J_2\mp d_{t_2})$& 1 & 3&1 & & & & &1\\
\hline
1 & & $E_{t_{3^\pm,\alpha}}  =  -\frac{1}{4}(J_1+3J_2\mp d_{t_2})$& 1 & 3 &-1 & & & & &-1\\
\hline
1 & 0 & $E_{t_{3^\pm,z}}  =  -\frac{1}{4}(J_1+3J_2\mp d_{t_2})$& 1 & 3 &-1 & & & & &-1\\
\hline
1 & & $E_{t_{4,\alpha}}=  -J_1$ & -1 & 3 &-1 & & & & &1\\
\hline
1 & 0 & $E_{t_{4,z}}=  -J_1$ & -1 & 3 & & &-1 & & 1&\\
\hline
1 & & $ E_{t_{5,\alpha}} =  -J_1$ & -1 & 3&1 & & & & &-1\\
\hline
1 & 0 & $ E_{t_{5,z}} =  -J_1$ & -1 & 3& & &1 & &-1 &\\
\hline
 1 & & $E_{t_{6,\alpha}}=\frac{1}{2}\left(J_1-3J_2\right) $ & -1 & 1&-1 &-1 &-1 &1 &1 &1\\
\hline
 1 & 0 & $E_{t_{6,z}}=\frac{1}{2}\left(J_1-3J_2\right) $ & -1 & 1&-1 &-1 &-1 &1 &1 &1\\
\hline
2 & & $ E_{q_{1,1^\pm}}= J_1$ & -1 & 3& &-1 & &1 & &\\
\hline
2 & & $ E_{q_{1,\alpha}}= J_1$ & -1 & 3& & &-1 & &1 &\\
\hline
2 & 0 & $ E_{q_{1,z}}= J_1$ & -1 & 3&-1 & & & & &1\\
\hline
2 & & $ E_{q_{2,1^\pm}} =  J_1$ & -1 & 3& &1 & &-1 & &\\
\hline
2 & &  $ E_{q_{2,\alpha}} =  J_1$ & -1 & 3& & &1 & &-1 &\\
\hline
2 & 0 & $ E_{q_{2,z}} =  J_1$ & -1 & 3&1 & & & & &-1\\
\hline
2 & & $E_{q_{3,1^\pm}}=0$& 1 & 3 & & & 1& &1 & \\
\hline
2 & & $E_{q_{3,\alpha}}=0$& 1 & 3 &1 & & & & &1 \\
\hline
2 &0 & $E_{q_{3,z}}=0$& 1 & 3 &1 & & & & & 1\\
\hline
2 & & $E_{q_{4,1^\pm}}=0$& 1 & 3& & & -1& &-1 & \\
\hline
2 & & $E_{q_{4,\alpha}}=0$& 1 & 3&-1 & & & & &-1\\
\hline
2 & 0 & $E_{q_{4,z}}=0$& 1 & 3&-1 & & & & &-1\\
\hline
2 & & $E_{q_{5,1^\pm}}=\frac{1}{2}\left(-J_1+3J_2\right)$& -1 & 1&1 &1 &1 &-1 &-1 &-1 \\
\hline
2 & & $E_{q_{5,\alpha}}=\frac{1}{2}\left(-J_1+3J_2\right)$& -1 & 1&1 &1 &1 &-1 &-1 &-1 \\
\hline
2 & 0 & $E_{q_{5,z}}=\frac{1}{2}\left(-J_1+3J_2\right)$& -1 & 1&1 &1 &1 &-1 &-1 &-1 \\
\hline
3 & & $ E_{h_{1^\pm}}=\frac{3}{2}\left(J_1+J_2\right)$& 1 & 1&1 &1 &1 &1 &1 &1 \\
\hline
3 & & $ E_{h_{2^\pm}}=\frac{3}{2}\left(J_1+J_2\right)$& 1 & 1&1 &1 &1 &1 &1 &1 \\
\hline
3 & & $ E_{h_{\alpha}}=\frac{3}{2}\left(J_1+J_2\right)$& 1 & 1&1 &1 &1 &1 &1 &1 \\
\hline
3 & 0 & $ E_{h_{z}}=\frac{3}{2}\left(J_1+J_2\right)$& 1 & 1&1 &1 &1 &1 &1 &1 \\
\hline
\end{tabular}
\end{center}
\end{table}
   \begin{table}[h]
 \def\arraystretch{0.5}
  \tabcolsep0.01pt\begin{tabular}{ll}
where,
 $d_s=\sqrt{13J_1^2+9J_2^2-18J_1J_2}$,
   $d_{t_1}=\sqrt{5J_1^2+9J_2^2-10J_1J_2}$ ,
  $ d_{t_2}=\sqrt{17J_1^2+9J_2^2-10J_1J_2}$.
  \end{tabular}
 \end{table} 
 All energy eigenvalues along with the corresponding eigenvalues 
of the group operations on the eigenstates are listed in the 
Table I. Here, the energy eigenstates 
of a definite energy value are constructed 
in such a fashion that they are the eigenstates of $S^z_{\rm T}$ only when 
the corresponding eigenvalue is zero. Otherwise they are expressed as a 
linear combinations of eigenstates of $S^z_{\rm T}$ with eigenvalues $\pm 1$, $\pm 2$, $\pm 3$, 
separately when $S_{\rm T}>0$. 
As a result, eigenvalue of $S^z_{\rm T}$ 
(second column of Table I) is not defined for every 
energy eigenstate. $\lambda_{{M}_n}$ does not always have definite value. 
The energy eigenstates are expressed in this 
way because of the fact that these forms are found useful to construct the 
spin operators in the plaquette operator theory as presented in the 
Appendix \ref{lambda}. However, energy eiegenstates with definite values of 
$S^z_{\rm T}$ are available for a more general 
Heisenberg hexagon in the article \cite{Moumita1}. 

To write down all the eigenstates following notations have been used.
\begin{equation}
\begin{aligned}
 & \ket{\psi^3_n}=T^{n-1}\ket{3}\left(n=1\right), \ket{3}=\ket{\uparrow\uparrow\uparrow\uparrow\uparrow\uparrow},\\
 &\ket{\psi^2_n}=T^{n-1}\ket{2}\left(n=1,2,3,4,5,6\right), \ket{2}=\ket{\uparrow\uparrow\uparrow\uparrow\uparrow\downarrow}, \\
 & \ket{\psi^1_n}_0=T^{n-1}\ket{1}_0\left(n=1,2,3,4,5,6\right), \ket{1}_0=\ket{\uparrow\uparrow\uparrow\downarrow\downarrow\uparrow},\\
 & \ket{\psi^1_n}_1=T^{n-1}\ket{1}_1\left(n=1,2,3,4,5,6\right), \ket{1}_1=\ket{\downarrow\uparrow\uparrow\uparrow\downarrow\uparrow}, \\
 & \ket{\psi^1_n}_2=T^{n-1}\ket{1}_2\left(n=1,2,3\right), \ket{1}_2=\ket{\downarrow\uparrow\uparrow\downarrow\uparrow\uparrow}, \\
  & \ket{\psi^0_n}_0=T^{n-1}\ket{0}_0\left(n=1,2,3,4,5,6\right), \ket{0}_0=\ket{\uparrow\uparrow\uparrow\downarrow\downarrow\downarrow}, \\
  & \ket{\psi^0_n}_1=T^{n-1}\ket{0}_1\left(n=1,2,3,4,5,6\right), \ket{0}_1=\ket{\uparrow\uparrow\downarrow\downarrow\uparrow\downarrow}, \\
  & \ket{\psi^0_n}_2=T^{n-1}\ket{0}_2\left(n=1,2,3,4,5,6\right), \ket{0}_2=\ket{\uparrow\downarrow\uparrow\downarrow\downarrow\uparrow}, \\
   & \ket{\psi^0_n}_3=T^{n-1}\ket{0}_3\left(n=1,2\right), \ket{0}_3=\ket{\uparrow\downarrow\uparrow\downarrow\uparrow\downarrow}, \\
    &  \ket{\psi^{-1}_n}_0=T^{n-1}\ket{-1}_0\left(n=1,2,3,4,5,6\right), \ket{-1}_0=\ket{\downarrow\downarrow\downarrow\uparrow\uparrow\downarrow},  \\
 & \ket{\psi^{-1}_n}_1=T^{n-1}\ket{-1}_1\left(n=1,2,3,4,5,6\right), \ket{-1}_1=\ket{\uparrow\downarrow\downarrow\downarrow\uparrow\downarrow}, \\
 & \ket{\psi^{-1}_n}_2=T^{n-1}\ket{-1}_2\left(n=1,2,3\right), \ket{-1}_2=\ket{\uparrow\downarrow\downarrow\uparrow\downarrow\downarrow}, \\
  & \ket{\psi^{-2}_n}=T^{n-1}\ket{-2}\left(n=1,2,3,4,5,6\right), \ket{-2}=\ket{\downarrow\downarrow\downarrow\downarrow\downarrow\uparrow}, \\
  & \ket{\psi^{-3}_n}=T^{n-1}\ket{-3}\left(n=1\right), \ket{-3}=\ket{\downarrow\downarrow\downarrow\downarrow\downarrow\downarrow}.
   \label{bra}
 \end{aligned}
\end{equation}
 Here $T$ is a unitary cyclic right shift operator. 
 $T\ket{abcdef}=\ket{fabcde} $ where 
$ \ket{abcdef}=\ket{a}\otimes\ket{b}\otimes\ket{c}\otimes\ket{d}\otimes\ket{e}\otimes\ket{f}$.
  All the energy eigenstates have been listed below.
  \small
   \begin{equation}
 \begin{aligned}
  \ket{s_{ 1^\pm}} &=\frac{1}{\mu_{s_{ 1^\pm}}\sqrt{12}}\Bigg(\sum\limits_{n=1,6} \left(-1\right)^{n-1} \Big(\sqrt{2}C_{s_{1^\pm,3}}
   \ket{\psi^0_n}_0+C_{s_{1^\pm,2}} 
    ( \ket{\psi^0_n}_1 + \ket{\psi^0_n}_2)\Big)+\sqrt{6}C_{s_{1^\pm,1}}\sum\limits_{n=1}^{2} \left(-1\right)^{n-1}  \ket{\psi^0_n}_3\Bigg) \\
\ket{s_2} &=\frac{1}{\sqrt{12}}\left(\sum\limits_{n=1}^{6}  \left( \ket{\psi^0_n}_2- \ket{\psi^0_n}_1\right)\right) \\
\ket{s_3}&=\frac{1}{2}\left(\sum\limits_{n=3,6} \left(-1\right)^{n}  \ket{\psi^0_n}_0+\sum\limits_{n=1,4}\left(-1\right)^{n} \left( \ket{\psi^0_n}_1+ \ket{\psi^0_n}_2\right)\right) 
   +\frac{1}{6}\sum\limits_{n=1}^{6}\left(-1\right)^{n-1}\left( \ket{\psi^0_n}_0+ \ket{\psi^0_n}_1+ \ket{\psi^0_n}_2\right)  \\
  \ket{s_4} &=\frac{1}{\sqrt{12}}\Bigg(\sum\limits_{n=1}^{2}  \ket{\psi^0_n}_0-\sum\limits_{n=4}^{5}  \ket{\psi^0_n}_0+\sum\limits_{n=5}      ^{6}       \left( \ket{\psi^0_n}_1+ \ket{\psi^0_n}_2\right)
        -\sum\limits_{n=2}^{3}  \left( \ket{\psi^0_n}_1+ \ket{\psi^0_n}_2\right)\Bigg) \nonumber\\  
\ket{t_{1^\pm,\alpha}}&= \frac{\lambda_{\alpha}}{\mu_{t^\alpha_{1^\pm}}\sqrt{12}}\Bigg(\sum\limits_{n=1}^{6}\Big(C_{t^\alpha_{1^\pm,1}}  
    \left( \ket{\psi^1_n}_0\mp\ket{\psi^{-1}_n}_0\right)+C_{t^\alpha_{1^\pm,2}} (\ket{\psi^1_n}_1
    \mp\ket{\psi^{-1}_n}_1)\Big) 
     +\sqrt{2}C_{t^\alpha_{1^\pm,3}}\sum\limits_{n=1}^{3}\left(\ket{\psi^1_n}_2\mp \ket{\psi^{-1}_n}_2\right) \Bigg) \\  
\ket{t_{1^\pm,z}}&= \frac{1}{\mu_{t^z_{1^\pm}}\sqrt{12}}\Bigg(\sum\limits_{n=1}^{6}\Big(\sqrt{2}C_{t^z_{1^\pm,3}} \ket{\psi^0_n}_0+C_{t^z_{1^\pm,1}} ( \ket{\psi^0_n}_1 
    + \ket{\psi^0_n}_2)\Big)+\sqrt{6}C_{t^z_{1^\pm,2}}\sum\limits_{n=1}^{2} \ket{\psi^0_n}_3 \Bigg) \\
        \ket{t_{2^\pm,\alpha}}&=\frac{\lambda_{\alpha}}{\mu_{t^\alpha_{2^\pm}}\sqrt{24}}\Bigg(3\sum\limits_{n=2,5} \Big(C_{t^\alpha_{2^\pm,1}}  \left( \ket{\psi^1_n}_0  \mp \ket{\psi^{-1}_n}_0 \right) 
  +  C_{t^\alpha_{2^\pm,2}} \left( \ket{\psi^1_n}_1  \mp \ket{\psi^{-1}_n}_1 \right)\Big)
  -\sum\limits_{n=1}^{6} \Big(C_{t^\alpha_{2^\pm,1}}  \left( \ket{\psi^1_n}_0  \mp \ket{\psi^{-1}_n}_0 \right) \\
 & + C_{t^\alpha_{2^\pm,2}} \left( \ket{\psi^1_n}_1  \mp \ket{\psi^{-1}_n}_1 \right)\Big)\Bigg) 
   +\frac{\lambda_{\alpha}}{\mu_{t^\alpha_{2^\pm}}\sqrt{12}}C_{t^\alpha_{2^\pm,3}}\left( 2\left(\ket{1}_2\mp\ket{-1}_2\right)-
  \sum\limits_{n=2}^{3} \left(\ket{\psi^1_n}_2\mp\ket{\psi^{-1}_n}_2 \right) \right) \\ 
   \ket{t_{2^\pm,z}}&= \frac{C_{t^z_{2^\pm,1}}}{\mu_{t^z_{2^\pm}}\sqrt{24}}\left( 3 \sum\limits_{n=1,4} \left( \ket{\psi^0_n}_1+ \ket{\psi^0_n}_2\right)
  -\sum\limits_{n=1}^{6}\left( \ket{\psi^0_n}_1+ \ket{\psi^0_n}_2\right)\right)
  +\frac{C_{t^z_{2^\pm,2}}}{\mu_{t^z_{2^\pm}}\sqrt{12}}\left(3\sum\limits_{n=3,6}\ket{\psi^0_n}_0-\sum\limits_{n=1}^{6} \ket{\psi^0_n}_0 \right)\\
 \end{aligned}
 \end{equation}  
 \begin{equation}
 \begin{aligned} 
  \ket{t_{3^\pm,\alpha}}&=\frac{\lambda_{\alpha}}{\sqrt{8}\mu_{t^\alpha_{3^\pm}}}\Bigg(\sum\limits_{n=1,4} \Bigg(C_{t^\alpha_{3^\pm,1}} \left( \ket{\psi^1_n}_0  \mp \ket{\psi^{-1}_n}_0 \right)
 + C_{t^\alpha_{3^\pm,2}}\left( \ket{\psi^1_n}_1  \mp \ket{\psi^{-1}_n}_1 \right)\Bigg)\\
&- \sum\limits_{n=3,6} \left( C_{t^\alpha_{3^\pm,1}} \left( \ket{\psi^1_n}_0  \mp \ket{\psi^{-1}_n}_0 \right)+   
 C_{t^\alpha_{3^\pm,2}}\left( \ket{\psi^1_n}_1  \mp \ket{\psi^{-1}_n}_1 \right)\right)\Bigg) 
 +\frac{\lambda_{\alpha}}{2\mu_{t^\alpha_{3^\pm}}} C_{t^\alpha_{3^\pm,3}}\sum\limits_{n=2}^{3}\left(-1\right)^{n-1} \left( \ket{\psi^1_n}_2  \mp \ket{\psi^{-1}_n}_2 \right) \\ 
 \ket{t_{3^\pm,z}}&= \frac{C_{t^z_{3^\pm,1}}}{\mu_{t^z_{3^\pm}}\sqrt{8}}\left(\sum\limits_{n=3,6}\left( \ket{\psi^0_n}_1+ \ket{\psi^0_n}_2\right)-
 \sum\limits_{n=2,5}\left( \ket{\psi^0_n}_1+ \ket{\psi^0_n}_2\right)\right) 
 +\frac{C_{t^z_{3^\pm,2}}}{2\mu_{t^z_{3^\pm}}}\left(\sum\limits_{n=2,5}  \ket{\psi^0_n}_0-\sum\limits_{n=1,4} \ket{\psi^0_n}_0\right)\\
 \ket{t_{4,\alpha}}&=\frac{\lambda_{\alpha}}{\sqrt{96}}\Bigg(\sum\limits_{n=1}^{6}\left(-1\right)^{n} \left( \ket{\psi^1_n}_0  \mp \ket{\psi^{-1}_n}_0 \right)
   + 3\sum\limits_{n=1,4}\left(-1\right)^{n-1} \left( \ket{\psi^1_n}_0 \mp\ket{\psi^{-1}_n}_0\right) \\
  &  + 3 \sum\limits_{n=2}^{3} \left( \ket{\psi^1_n}_1 \mp \ket{\psi^{-1}_n}_1\right)
    -3 \sum\limits_{n=5}^{6} \left(  \ket{\psi^1_n}_1  \mp\ket{\psi^{-1}_n}_1 \right) \Bigg)  \\
     \ket{t_{4,z}}&= \frac{1}{\sqrt{24}}\Bigg(\sum\limits_{n=1}^{6}\left(-1\right)^{n}  \left( \ket{\psi^0_n}_1- 
     \ket{\psi^0_n}_2\right)
     +3\sum\limits_{n=3,6}\left(-1\right)^{n-1}\left( \ket{\psi^0_n}_1- \ket{\psi^0_n}_2\right) \Bigg)   \\
           \ket{t_{5,\alpha}}&=\frac{\lambda_{\alpha}}{4\sqrt{2}}\Bigg(\sum\limits_{n=2}^{3} \left( \ket{\psi^1_n}_0  \mp \ket{\psi^{-1}_n}_0 \right) -\sum\limits_{n=5}^{6}  
  \left( \ket{\psi^1_n}_0  \mp \ket{\psi^{-1}_n}_0 \right)
  + \sum\limits_{n=1}^{6}\left(-1\right)^{n-1}  \left( \ket{\psi^1_n}_1  \mp \ket{\psi^{-1}_n}_1 \right) \\
 & + 3\sum\limits_{n=1,4}\left(-1\right)^{n}  \left( \ket{\psi^1_n}_1  \mp \ket{\psi^{-1}_n}_1 \right)   \Bigg)     \\
 \ket{t_{5,z}}&= \frac{1}{\sqrt{8}}\Bigg(\sum\limits_{n=1,2}\left(-1\right)^{n}  \left( \ket{\psi^0_n}_2- \ket{\psi^0_n}_1\right)
  +\sum\limits_{n=4,5}\left(-1\right)^{n-1}\left( \ket{\psi^0_n}_1- \ket{\psi^0_n}_2\right) \Bigg)   \nonumber  \\
   \ket{t_{6,\alpha}}&=\frac{\lambda_{\alpha}}{\sqrt{12}}\sum\limits_{n=1}^{6}\left(-1\right)^{n}   \left(\ket{\psi^1_n}_0\mp \ket{\psi^{-1}_n}_0 \right)    \\
  \ket{t_{6,z}}&=\frac{1}{\sqrt{12}}\left(\sum\limits_{n=1}^{6}\left(-1\right)^{n-1} \left( \ket{\psi^0_n}_1 -\ket{\psi^0_n}_2\right) \right) \\
   \ket{q_{1,1^\pm}} &=\frac{1}{2\sqrt{2}} \left(\sum\limits_{n=1}^{2} \left( \ket{\psi^2_n}\pm \ket{\psi^{-2}_n}\right)
 -\sum\limits_{n=4}^{5}\left( \ket{\psi^2_n}\pm \ket{\psi^{-2}_n}\right)\right) \nonumber\\
  \ket{q_{1,\alpha}}&=\frac{\lambda_{\alpha}}{4\sqrt{2}}\Bigg(\sum\limits_{n=1}^{6}\left(-1\right)^{n-1}\left(\ket{\psi^1_n}_0\mp \ket{\psi^{-1}_n}_0\right) 
  +3 \sum\limits_{n=1,4} \left(-1\right)^{n} \left(\ket{\psi^1_n}_0\mp \ket{\psi^{-1}_n}_0\right) \\
 &  +\sum\limits_{n=2}^{3}  \left(\ket{\psi^1_n}_1\mp \ket{\psi^{-1}_n}_1\right) - 
  \sum\limits_{n=5}^{6} \left(\ket{\psi^1_n}_1\mp \ket{\psi^{-1}_n}_1\right)  \Bigg) \\
 \ket{q_{1,z}}&=  \frac{1}{\sqrt{24}}\Bigg(2 \sum\limits_{n=1}^{2}  \ket{\psi^0_n}_0-2\sum\limits_{n=4}^{5}  \ket{\psi^0_n}_0+
   \sum\limits_{n=2}^{3}  \left( \ket{\psi^0_n}_1 +  \ket{\psi^0_n}_2\right)
   -\sum\limits_{n=5}^{6} \left( \ket{\psi^0_n}_1 +  \ket{\psi^0_n}_2\right)\Bigg)\\   
    \ket{q_{2,1^\pm}} &=\frac{1}{\sqrt{24}}\Bigg(\sum\limits_{n=1}^{6}\left(-1\right)^{n-1} 
\left( \ket{\psi^2_n}\pm \ket{\psi^{-2}_n}\right)
+3 \sum\limits_{n=3,6} \left(-1\right)^{n} \left( \ket{\psi^2_n}\pm \ket{\psi^{-2}_n}\right)\Bigg) \\
\ket{q_{2,\alpha}}&=\frac{\lambda_{\alpha}}{\sqrt{96}}\Bigg(3\sum\limits_{n=2}^{3} \left(\ket{\psi^1_n}_0\mp \ket{\psi^{-1}_n}_0\right)
   - 3\sum\limits_{n=5}^{6} \left(\ket{\psi^1_n}_0\mp \ket{\psi^{-1}_n}_0\right)
   +\sum\limits_{n=1}^{6}\left(-1\right)^{n}  \left(\ket{\psi^1_n}_1\mp \ket{\psi^{-1}_n}_1\right)\\
   &+3 \sum\limits_{n=1,4} \left(-1\right)^{n-1} \left(\ket{\psi^1_n}_1\mp \ket{\psi^{-1}_n}_1\right)   \Bigg) \\
   \ket{q_{2,z}}&=   \frac{1}{\sqrt{72}}\Bigg(6\sum\limits_{n=3,6}\left(-1\right)^{n}  \ket{\psi^0_n}_0 
   +3\sum\limits_{n=1,4}\left(-1\right)^{n-1}  \left( \ket{\psi^0_n}_1 +  \ket{\psi^0_n}_2\right)
  +\sum\limits_{n=1}^{6}\left(-1\right)^{n} \left( \ket{\psi^0_n}_1 +  \ket{\psi^0_n}_2-2 \ket{\psi^0_n}_0\right)\Bigg) \\
   \ket{q_{3,1^\pm}} &=\frac{1}{\sqrt{24}}\left(3\sum\limits_{n=1,4}  \left( \ket{\psi^2_n}\pm \ket{\psi^{-2}_n}\right)
  - \sum\limits_{n=1}^{6} \left( \ket{\psi^2_n}\pm \ket{\psi^{-2}_n}\right)\right) \\
 \ket{q_{3,\alpha}}&= \frac{\lambda_{\alpha}}{\sqrt{96}}\Bigg(3\sum\limits_{n=2,5}  \left(  \left(\ket{\psi^1_n}_0\mp \ket{\psi^{-1}_n}_0\right)  
    +  \left(\ket{\psi^1_n}_1\mp \ket{\psi^{-1}_n}_1\right) \right)
    -\sum\limits_{n=1}^{6} \Big(  \left(\ket{\psi^1_n}_0\mp \ket{\psi^{-1}_n}_0\right) \\
   & +  \left(\ket{\psi^1_n}_1\mp \ket{\psi^{-1}_n}_1\right) \Big) 
    +2 \sum\limits_{n=2}^{3}  \left(\ket{\psi^1_n}_2\mp\ket{\psi^{-1}_n}_2\right) -4\left(\ket{1}_2 \mp \ket{-1}_2\right)  \Bigg) \\
  \end{aligned}
     \end{equation}     
    \small 
 \begin{equation}
 \begin{aligned}         
    \label{vector}  
   \ket{q_{3,z}}&= \frac{1}{\sqrt{8}}\left(\sum\limits_{n=2,5}  \left( \ket{\psi^0_n}_2
     -  \ket{\psi^0_n}_1\right)+\sum\limits_{n=3,6}\left( \ket{\psi^0_n}_1 -  \ket{\psi^0_n}_2\right) \right) \\
\ket{q_{4,1^\pm}} &=\frac{1}{2\sqrt{2}}\left(\sum\limits_{n=3,6} \left( \ket{\psi^2_n}\pm \ket{\psi^{-2}_n}\right)-
 \sum\limits_{n=2,5} \left( \ket{\psi^2_n}\pm \ket{\psi^{-2}_n}\right)\right) \\
         \ket{q_{4,\alpha}}&= \frac{\lambda_{\alpha}}{4\sqrt{2}}\Bigg(\sum\limits_{n=1,4} \left(  \left(\ket{\psi^1_n}_0\mp \ket{\psi^{-1}_n}_0\right)  +
   \left(\ket{\psi^1_n}_1\mp \ket{\psi^{-1}_n}_1\right) \right)
   -\sum\limits_{n=3,6} \Big(  \left(\ket{\psi^1_n}_0\mp \ket{\psi^{-1}_n}_0\right) \\
  &+   \left(\ket{\psi^1_n}_1\mp \ket{\psi^{-1}_n}_1\right) \Big) 
  +2 \sum\limits_{n=2}^{3}\left(-1\right)^{n}   \left(\ket{\psi^1_n}_2\pm \ket{\psi^{-1}_n}_2\right)  \Bigg) \\
    \ket{q_{4,z}}&= \frac{1}{\sqrt{24}}\left(3\sum\limits_{n=1,4}  \left( \ket{\psi^0_n}_1 -  \ket{\psi^0_n}_2\right)
       +\sum\limits_{n=1}^{6}\left( \ket{\psi^0_n}_2 -  \ket{\psi^0_n}_1\right) \right)\\
       \ket{q_{5,1^\pm}} &= \frac{1}{\sqrt{12}}\left(\sum\limits_{n=1}^{6}\left(-1\right)^{n-1} \left( \ket{\psi^2_n}\pm \ket{\psi^{-2}_n}\right)\right) \nonumber\\
        \ket{q_{5,\alpha}}&=\frac{\lambda_{\alpha}}{\sqrt{12}}\left(\sum\limits_{n=1}^{6}\left(-1\right)^{n-1} 
 \left(\ket{\psi^1_n}_1\mp \ket{\psi^{-1}_n}_1\right) \right) \\
  \ket{q_{5,z}}&= \frac{1}{6}\Bigg(\sum\limits_{n=1}^{6}\left(-1\right)^{n-1}  \left( \ket{\psi^0_n}_0+ \ket{\psi^0_n}_1 
         +  \ket{\psi^0_n}_2\right)
         +3\sum\limits_{n=1}^{2}\left(-1\right)^{n-1} \ket{\psi^0_n}_3 \Bigg) \\
   \ket{h_{1^\pm}}&=\frac{1}{\sqrt{2}}\left(\ket{\psi^3_n}\pm\ket{\psi^{-3}_n}\right) \\  
        \ket{h_{2^\pm}}&= \frac{1}{\sqrt{12}}\sum\limits_{n=1}^{6}\left(  \ket{\psi^2_n}\pm  \ket{\psi^{-2}_n}\right)\\
    \ket{h_\alpha}&= \frac{\lambda_{\alpha}}{\sqrt{30}}\Bigg(\sum\limits_{n=1}^{6} \left( \left( \ket{\psi^1_n}_0 \mp \ket{\psi^{-1}_n}_0\right) +  
     \left( \ket{\psi^1_n}_1 \mp \ket{\psi^{-1}_n}_1\right)\right) 
     +  \sum\limits_{n=1}^{3}  \left(\ket{\psi^1_n}_2\mp \ket{\psi^{-1}_n}_2\right) \Bigg) \\
      \ket{h_z}&= \frac{1}{\sqrt{20}}\left(\sum\limits_{n=1}^{6} \left(  \ket{\psi^0_n}_0 +  \ket{\psi^0_n}_1 + \ket{\psi^0_n}_2 \right) + 
      \sum\limits_{n=1}^{2}  \ket{\psi^0_n}_3 \right)\nonumber 
  \end{aligned}
     \end{equation} 
     \normalsize
 where the upper and lower signs respectively refer to $\alpha = x$
and $y$, $\lambda_x =-1$ and $\lambda_y = i$.
\small
 \begin{equation}
 \begin{aligned}
    &  C_{s_{1^\pm,1}}=\frac{-\sqrt{6}J_1}
   {\left(-J_1+3J_2\mp d_s\right)}, 
  C_{s_{1^\pm,2}}=1, 
   C_{s_{1^\pm,3}}=\frac{-\sqrt{2}\left(2J_2-J_1\right)}
   {\left(3J_1-J_2\mp d_s\right)} \\
  &C_{t^\alpha_{1^\pm,1}}=\frac{2J_1\left(J_1-J_2\mp d_{t_1}\right)-8J_1J_2}{8J_2^2-\left(3J_1+J_2\mp d_{t_1}\right)\left(J_1-J_2\mp d_{t_1}\right)},  
   C_{t^\alpha_{1^\pm,2}}=1, \\
 &  C_{t^\alpha_{1^\pm,3}}=\frac{2\sqrt{2}\left(2J_1J_2-J_1\left(3J_1+J_2\mp d_{t_a}\right)\right)}
  {\left(3J_1+J_2\mp d_{t_a}\right)\left(J_1-J_2\mp d_{t_a}\right)-8J_2^2},\\
      & C_{t^z_{1^\pm,1}}=\frac{\sqrt{6}J_1\left(3J_1-J_2\mp d_{t_1}\right)}
    {2\left(2J_2+J_1\right)^2-\left(3J_1+J_2\mp d_{t_1}\right)\left(3J_1-J_2\mp d_{t_1}\right)}, 
    C_{t^z_{1^\pm,2}}=1, \\
   & C_{t^z_{1^\pm,3}}= \frac{-2\sqrt{3}J_1\left(2J_2+J_1\right)}
    {2\left(2J_2+J_1\right)^2-\left(3J_1+J_2\mp d_{t_1}\right)\left(3J_1-J_2\mp d_{t_1}\right) }, \\
   &C_{t^\alpha_{2^\pm,1}}\!\!\!=\!\!C_{t^\alpha_{3^\pm,1}}\!\!\!=\!\!\frac{8J_1J_2+2J_1\left(-J_1+J_2\mp d_{t_2}\right)}
   {\left(3J_1-J_2\mp d_{t_2}\right)\left(-J_1+J_2\mp d_{t_2}\right)-8J_2^2}, 
    C_{t^\alpha_{2^\pm,2}}\!\!\!=\!\!C_{t^\alpha_{3^\pm,2}}\!\!\!=\!\!1,\\
    & C_{t^\alpha_{2^\pm,3}}= C_{t^\alpha_{3^\pm,3}}=\frac{2\sqrt{2}\left(J_1\left(3J_1-J_2\mp d_{t_2}\right)+2J_2J_1\right)}
     {\left(3J_1-J_2\mp d_{t_2}\right)\left(-J_1+J_2\mp d_{t_2}\right)-8J_2^2}, \\
 & C_{t^z_{2^\pm,1}}=C_{t^z_{3^\pm,1}}=\frac{2\sqrt{2}\left(J_1-J_2\right)}{3J_1+J_2\pm d_{t_2}}, 
     C_{t^z_{2^\pm,2}}=C_{t^z_{3^\pm,2}}=1, \\
  &\mu_{s_{ 1^\pm}}=\sqrt{C^2_{s_{1^\pm,1}}+C^2_{s_{1^\pm,2}}+C^2_{s_{1^\pm,3}}},  
   \mu_{t^\alpha_{1^\pm}}=\sqrt{C^2_{t^\alpha_{1^\pm,1}}+C^2_{t^\alpha_{1^\pm,2}}+C^2_{t^\alpha_{1^\pm,3}}},\\
  & \mu_{t^z_{1^\pm}}=\sqrt{C^2_{t^z_{1^\pm,1}}+C^2_{t^z_{1^\pm,2}}+C^2_{t^z_{1^\pm,3}}}  , 
  \mu_{t^z_{2^\pm}}=\mu_{t^z_{3^\pm}}=\sqrt{C^2_{t^z_{2^\pm,1}}+C^2_{t^z_{2^\pm,2}}},\\
  & \mu_{t^\alpha_{2^\pm}}=\mu_{t^\alpha_{3^\pm}}=\sqrt{C^2_{t^\alpha_{2^\pm,1}}+C^2_{t^\alpha_{2^\pm,2}}+C^2_{t^\alpha_{2^\pm,3}}}. \nonumber\\
    \end{aligned}
  \end{equation}
    \normalsize
\section{DETAILS OF PLAQUETTE OPERATOR THEORY}
  \label{lambda}
The six different spin operators, $S^n_\alpha$, are expressed 
in terms of all singlet and triplet boson operators. 
\scriptsize
 \begin{equation}
 \begin{aligned}
 S^n_\alpha&=\frac{\left(-1\right)^{n-1}}{2\mu_{s_{ 1^\pm}}\mu_{t^z_{1^\pm}}}\left(C_{s_{1^\pm,1}}C_{t^z_{1^\pm,2}}+\frac{C_{s_{1^\pm,2}}C_{t^z_{1^\pm,1}}}{3}
 +\frac{C_{s_{1^\pm,3}}C_{t^z_{1^\pm,3}}}{3}\right)\left(t^\dagger_{1^\pm,\alpha}s_{ 1^\pm}+s^\dagger_{ 1^\pm}t_{1^\pm,\alpha}\right)\\
 &+ 
 \frac{\left(-1\right)^{n-1}\lambda_{n,1}}{3\sqrt{2}\mu_{s_{ 1^\pm}}\mu_{t^z_{2^\pm}}}\left(C_{s_{1^\pm,2}}C_{t^z_{2^\pm,1}}-2C_{s_{1^\pm,3}}C_{t^z_{2^\pm,2}}\right)
 \left(t^\dagger_{2^\pm,\alpha}s_{ 1^\pm}+s^\dagger_{ 1^\pm}t_{2^\pm,\alpha}\right)\\
 &+\frac{\left(-1\right)^{n-1}\lambda_{n,3}}{2\sqrt{6}\mu_{s_{ 1^\pm}}\mu_{t^z_{3^\pm}}}\left(C_{s_{1^\pm,2}}C_{t^z_{3^\pm,1}}-2C_{s_{1^\pm,3}}C_{t^z_{3^\pm,2}}\right)
 \left(t^\dagger_{3^\pm,\alpha}s_{ 1^\pm}+s^\dagger_{ 1^\pm}t_{3^\pm,\alpha}\right)\\
& +\frac{\lambda_{n,4}}{2\sqrt{2}\mu_{s_{ 1^\pm}}}C_{s_{1^\pm,2}}\left(t^\dagger_{4,\alpha}s_{ 1^\pm}+s^\dagger_{ 1^\pm}t_{4,\alpha}\right)-
 \frac{\lambda_{n,5}}{2\sqrt{6}\mu_{s_{ 1^\pm}}}C_{s_{1^\pm,2}}\left(t^\dagger_{5,\alpha} s_{ 1^\pm}+ s^\dagger_{ 1^\pm}t_{5,\alpha}\right)\\
 &+ 
 \frac{\lambda_{n,3}}{2\sqrt{2}\mu_{t^z_{2^\pm}}}C_{t^z_{2^\pm,1}}\left(t^\dagger_{2^\pm,\alpha}s_{2}+s^\dagger_{2}t_{2^\pm,\alpha}\right)
-\frac{\lambda_{n,1}}{\sqrt{6}\mu_{t^z_{3^\pm}}}C_{t^z_{3^\pm,1}} \left(t^\dagger_{3^\pm,\alpha}s_{2}+s^\dagger_{2}t_{3^\pm,\alpha}\right)\\
&+\frac{\left(-1\right)^{n-1}\lambda_{n,5}}{6\sqrt{2}}\left(t^\dagger_{4,\alpha}s_{2}+s^\dagger_{2}t_{4,\alpha}\right)+
 \frac{\left(-1\right)^{n-1}\lambda_{n,4}}{2\sqrt{6}}\left(t^\dagger_{5,\alpha} s_{2}+ s^\dagger_{2}t_{5,\alpha}\right)+
 \frac{\left(-1\right)^{n}}{6}\left(t^\dagger_{6,\alpha}s_2+ s^\dagger_{2}t_{6,\alpha}\right)\\
 &+\frac{(-1)^{n-1}\lambda_{n,1}}{3\sqrt{3}\mu_{t^z_{1^\pm}}}\left(\sqrt{2}C_{t^z_{1^\pm,3}}\!\!\!-C_{t^z_{1^\pm,1}}\right)\left(t^\dagger_{1^\pm,\alpha}s_{3}+s^\dagger_{3}t_{1^\pm,\alpha}\right)
+\frac{(-1)^{n}}{6\sqrt{3}\mu_{t^z_{2^\pm}}}\left(\sqrt{2}\lambda_{n,11}C_{t^z_{2^\pm,1}}\!\!\!-\lambda_{n,10}C_{t^z_{2^\pm,2}}\right)\left(t^\dagger_{2^\pm,\alpha}s_{3}+s^\dagger_{3}t_{2^\pm,\alpha}\right)  \\
  &+\frac{(-1)^{n-1}\lambda_{n,3}}{6\sqrt{2}\mu_{t^z_{3^\pm}}}\left(C_{t^z_{3^\pm,1}}-\sqrt{2}C_{t^z_{3^\pm,2}}\right) \left(t^\dagger_{3^\pm,\alpha}s_{3}+s^\dagger_{3}t_{3^\pm,\alpha}\right)
   +\frac{\lambda_{n,6}}{2\sqrt{6}}\left(t^\dagger_{4,\alpha}s_{3}+s^\dagger_{3}t_{4,\alpha}\right)
 -\frac{\lambda_{n,7}}{6\sqrt{2}}\left(t^\dagger_{5,\alpha}s_{3}+s^\dagger_{3}t_{5,\alpha}\right)\\
 &+\frac{\lambda_{n,3}}{2\sqrt{3}}\left(t^\dagger_{6,\alpha}s_{3}+s^\dagger_{3}t_{6,\alpha}\right) 
 +\frac{\lambda_{n,3}(-1)^{n-1}}{6\mu_{t^z_{1^\pm}}}\left(\sqrt{2}C_{t^z_{1^\pm,3}}-C_{t^z_{1^\pm,1}}\right)\left(t^\dagger_{1^\pm,\alpha}s_{4}+s^\dagger_{4}t_{1^\pm,\alpha}\right)\\
&+\frac{(-1)^{n-1}\lambda_{n,3}}{6\sqrt{2}\mu_{t^z_{2^\pm}}}\left(C_{t^z_{2^\pm,1}}-\sqrt{2}C_{t^z_{2^\pm,2}}\right)\left(t^\dagger_{2^\pm,\alpha}s_{4}+s^\dagger_{4}t_{2^\pm,\alpha}\right)  
  +\frac{(-1)^{n}}{2\sqrt{3}\mu_{t^z_{3^\pm}}}\left(\frac{\lambda_{n,2}}{\sqrt{2}}C_{t^z_{3^\pm,1}}+\lambda_{n,14}C_{t^z_{3^\pm,2}}\right) \left(t^\dagger_{3^\pm,\alpha}s_{4}+s^\dagger_{4}t_{3^\pm,\alpha}\right)\\
 &  -\frac{\lambda_{n,7}}{6\sqrt{2}}\left(t^\dagger_{4,\alpha}s_{4}+s^\dagger_{4}t_{4,\alpha}\right)
 -\frac{\lambda_{n,6}}{2\sqrt{6}}\left(t^\dagger_{5,\alpha}s_{4}+s^\dagger_{4}t_{5,\alpha}\right)
 -\frac{\lambda_{n,1}}{3}\left(t^\dagger_{6,\alpha}s_{4}+s^\dagger_{4}t_{6,\alpha}\right)\\
 &- \frac{i}{6\mu_{t^x_{1^\pm}}\mu_{t^x_{1^\pm}}}\epsilon^{\alpha\beta\gamma}\left(C_{t^x_{1^\pm,1}}C_{t^x_{1^\pm,1}}+
 C_{t^x_{1^\pm,2}}C_{t^x_{1^\pm,2}}+C_{t^x_{1^\pm,3}}C_{t^x_{1^\pm,3}}\right)
 \left(t^\dagger_{1^\pm,\beta}t_{1^\pm,\gamma} +t^\dagger_{1^\pm,\beta}t_{1^\pm,\gamma}\right) \\
  &- \frac{i\lambda_{n,1}}{3\sqrt{2}\mu_{t^x_{1^\pm}}\mu_{t^x_{2^\pm}}}\epsilon^{\alpha\beta\gamma}\left(C_{t^x_{1^\pm,1}}C_{t^x_{2^\pm,1}}+
 C_{t^x_{1^\pm,2}}C_{t^x_{2^\pm,2}}-2C_{t^x_{1^\pm,3}}C_{t^x_{2^\pm,3}}\right)
 \left(t^\dagger_{1^\pm,\beta}t_{2^\pm,\gamma} +t^\dagger_{2^\pm,\beta}t_{1^\pm,\gamma}\right) \\
  &- \frac{i\lambda_{n,3}}{2\sqrt{6}\mu_{t^x_{1^\pm}}\mu_{t^x_{3^\pm}}}\epsilon^{\alpha\beta\gamma}\left(C_{t^x_{1^\pm,1}}C_{t^x_{3^\pm,1}}+
 C_{t^x_{1^\pm,2}}C_{t^x_{3^\pm,2}}-2C_{t^x_{1^\pm,3}}C_{t^x_{3^\pm,3}}\right)
 \left(t^\dagger_{1^\pm,\beta}t_{3^\pm,\gamma} +t^\dagger_{3^\pm,\beta}t_{1^\pm,\gamma}\right) \\
  &- \frac{i(-1)^{n-1}\lambda_{n,4}}{4\sqrt{2}\mu_{t^x_{1^\pm}}}\epsilon^{\alpha\beta\gamma}\left(C_{t^x_{1^\pm,1}}-
 C_{t^x_{1^\pm,2}}\right)
 \left(t^\dagger_{1^\pm,\beta}t_{4,\gamma} +t^\dagger_{4,\beta}t_{1^\pm,\gamma}\right) \\
  &- \frac{i(-1)^{n}\lambda_{n,5}}{4\sqrt{6}\mu_{t^x_{1^\pm}}}\epsilon^{\alpha\beta\gamma}\left(C_{t^x_{1^\pm,1}}-
 C_{t^x_{1^\pm,2}}\right)
 \left(t^\dagger_{1^\pm,\beta}t_{5,\gamma} +t^\dagger_{5,\beta}t_{1^\pm,\gamma}\right) \\
  &-\frac{i}{6\mu_{t^x_{2^\pm}}\mu_{t^x_{2^\pm}}}\epsilon^{\alpha\beta\gamma}\left(\lambda_{n,11}(C_{t^x_{2^\pm,1}}C_{t^x_{2^\pm,1}}+
 C_{t^x_{2^\pm,2}}C_{t^x_{2^\pm,2}})-\lambda_{n,10}C_{t^x_{2^\pm,3}}C_{t^x_{2^\pm,3}}\right)
 \left(t^\dagger_{2^\pm,\beta}t_{2^\pm,\gamma} +t^\dagger_{2^\pm,\beta}t_{2^\pm,\gamma}\right) \\
  &+ \frac{i\lambda_{n,3}}{4\sqrt{3}\mu_{t^x_{2^\pm}}\mu_{t^x_{3^\pm}}}\epsilon^{\alpha\beta\gamma}\left(C_{t^x_{2^\pm,1}}C_{t^x_{3^\pm,1}}+
 C_{t^x_{2^\pm,2}}C_{t^x_{3^\pm,2}}-2C_{t^x_{3^\pm,3}}C_{t^x_{3^\pm,3}}\right)
 \left(t^\dagger_{2^\pm,\beta}t_{3^\pm,\gamma} +t^\dagger_{3^\pm,\beta}t_{2^\pm,\gamma}\right) \\
   &-\frac{i(-1)^{n}}{8\mu_{t^x_{2^\pm}}}\epsilon^{\alpha\beta\gamma}
  \left(\lambda_{n,6}C_{t^x_{2^\pm,1}}-\lambda_{n,12}C_{t^x_{2^\pm,2}}\right)
  \left(t^\dagger_{2^\pm,\beta}t_{4,\gamma} +t^\dagger_{4,\beta}t_{2^\pm,\gamma}\right)\\
    &-\frac{i(-1)^{n-1}}{8\sqrt{3}\mu_{t^x_{2^\pm}}}\epsilon^{\alpha\beta\gamma}
  \left(\lambda_{n,7}C_{t^x_{2^\pm ,1}}-\lambda_{n,15}C_{t^x_{2^\pm,2}}\right)
  \left(t^\dagger_{2^\pm,\beta}t_{5,\gamma} +t^\dagger_{5,\beta}t_{2^\pm,\gamma}\right)\\
   &-\frac{i(-1)^{n-1}\lambda_{n,3}}{2\sqrt{2}\mu_{t^x_{2^\pm}}}\epsilon^{\alpha\beta\gamma}C_{t^x_{2^\pm,1}}
 \left(t^\dagger_{6,\beta}t_{2^\pm,\gamma} +t^\dagger_{2^\pm,\beta}t_{6,\gamma}\right) \\
 &- \frac{i}{2\mu_{t^x_{3^\pm}}\mu_{t^x_{3^\pm}}}\epsilon^{\alpha\beta\gamma}\left(\frac{\lambda_{n,2}}{2}(C_{t^x_{3^\pm,1}}C_{t^x_{3^\pm,1}}+
 C_{t^x_{3^\pm,2}}C_{t^x_{3^\pm,2}})+\lambda_{n,14}C_{t^x_{3^\pm,3}}C_{t^x_{3^\pm,3}}\right)
 \left(t^\dagger_{3^\pm,\beta}t_{3^\pm,\gamma} +t^\dagger_{3^\pm,\beta}t_{3^\pm,\gamma}\right) \\ 
    &-\frac{i(-1)^{n-1}}{8\sqrt{3}\mu_{t^x_{3^\pm}}}\epsilon^{\alpha\beta\gamma}
  \left(\lambda_{n,7}C_{t^x_{3^\pm,1}}+\lambda_{n,13}C_{t^x_{3^\pm,2}}\right)
  \left(t^\dagger_{3^\pm,\beta}t_{4,\gamma} +t^\dagger_{4,\beta}t_{3^\pm,\gamma}\right)\\
      &-\frac{i(-1)^{n-1}}{8\mu_{t^x_{3^\pm}}}\epsilon^{\alpha\beta\gamma}
  \left(\lambda_{n,6}C_{t^x_{3^\pm,1}}+\lambda_{n,16}C_{t^x_{3^\pm,2}}\right)
  \left(t^\dagger_{3^\pm,\beta}t_{5,\gamma} +t^\dagger_{5,\beta}t_{3^\pm,\gamma}\right)\\
    &- \frac{i(-1)^{n}\lambda_{n,1}}{\sqrt{6}\mu_{t^x_{3^\pm}}}\epsilon^{\alpha\beta\gamma}C_{t^x_{3^\pm,1}} 
 \left(t^\dagger_{3^\pm,\beta}t_{6,\gamma} +t^\dagger_{6,\beta}t_{3^\pm,\gamma}\right)
  - \frac{5i\lambda_{n,8}}{12}\epsilon^{\alpha\beta\gamma} t^\dagger_{4,\beta}t_{4,\gamma} 
 +\frac{i\lambda_{n,4}}{8\sqrt{3}}\epsilon^{\alpha\beta\gamma}
 \left(t^\dagger_{4,\beta}t_{5,\gamma} +t^\dagger_{5,\beta}t_{4,\gamma}\right) \\
 &-\frac{i\lambda_{n,5}}{12\sqrt{2}}\epsilon^{\alpha\beta\gamma}
 \left(t^\dagger_{4,\beta}t_{6,\gamma} +t^\dagger_{6,\beta}t_{4,\gamma}\right)
     - \frac{i\lambda_{n,9}}{4}\epsilon^{\alpha\beta\gamma} t^\dagger_{5,\beta}t_{5,\gamma} 
   -\frac{i\lambda_{n,4}}{4\sqrt{6}}\epsilon^{\alpha\beta\gamma}
 \left(t^\dagger_{5,\beta}t_{6,\gamma} +t^\dagger_{6,\beta}t_{5,\gamma}\right)
 -\frac{i}{3}\epsilon^{\alpha\beta\gamma}
 t^\dagger_{6,\beta}t_{6,\gamma}. 
 \label{operator}
    \end{aligned}
  \end{equation}
 \normalsize
 The values of $\lambda_{n,k}$ are given below, where $k=1,2,\cdots,19$. 
 Here, $\alpha,\beta,\gamma=x,y,z$ and $  \epsilon^ {\alpha\beta\gamma}$ is 
 the completely antisymmetric tensor 
 with $ \epsilon^ {xyz}=1$.
Summation convention over repeated indices is implied.
   \begin{table}[h]
\def\arraystretch{0.78}
  \tabcolsep4pt\begin{tabular}{|c|c|c|c||c|c|c|c|}
\hline
 $\!\boldsymbol\lambda_{n,k}\!\!$ & $\!\boldsymbol n\!=\!1,4\!\!$ & $\!\boldsymbol n\!=\!2,5\!\!$ & $\!\boldsymbol n\!=\!3,6\!$& $\!\!\boldsymbol\lambda_{n,k}\!\!$ 
 & $\!\boldsymbol n\!=\!1,4\!\!$ & $\!\boldsymbol n\!=\!2,5\!\!$ & $\!\boldsymbol n\!=\!3,6\!$\\
  \hline
  $\lambda_{n,1}$ & 1 & $-\frac{1}{2}$ & $-\frac{1}{2}$ &  $\lambda_{n,9}$ &1 & 1 & 2\\
  \hline
 $\lambda_{n,2}$ &0 & $1$ & $1$  &  $\lambda_{n,10}$ &1 & -2 & -2\\
  \hline
   $\lambda_{n,3}$ &0 & -1 & 1 &  $\lambda_{n,11}$ &2 & $\frac{1}{2}$ & $\frac{1}{2}$\\
  \hline
   $\lambda_{n,4}$ &1 & -1 & 0 &   $\lambda_{n,12}$ &1 & 2 & 3\\
  \hline
   $\lambda_{n,5}$ &1 & 1 & -2 & $\lambda_{n,13}$ &3 & 0 & 3  \\
  \hline
   $\lambda_{n,6}$ &1 & 0 & -1 & $\lambda_{n,14}$ &1 & 0 & 0 \\
  \hline
   $\lambda_{n,7}$ &1 & -2 & 1 & $\lambda_{n,15}$ &1 & 4 & 1 \\
  \hline
   $\lambda_{n,8}$ &1 & 1 & $\frac{2}{5}$ & $\lambda_{n,16}$ &3 & 2 & 1 \\
      \hline
 \end{tabular}
 \end{table} 

The expressions of the coefficients $A^n_{\eta}$, 
$B^n_{\eta}$ and $D^n_{\eta\xi}$ are given below. 
  \begin{equation}
 \begin{aligned}
 & A^n_{1}=\frac{\left(-1\right)^{n-1}}{2\mu_{s_{1^\Minus}}\mu_{t^z_{1^\Minus}}}\left(C_{s_{{1^\Minus},1}}C_{t^z_{{ 1^\Minus},2}}+\frac{C_{s_{{1^\Minus},2}}C_{t^z_{{ 1^\Minus},1}}}{3}
 +\frac{C_{s_{{1^\Minus},3}}C_{t^z_{{ 1^\Minus},3}}}{3}\right), \\
 & A^n_{2}= \frac{\left(-1\right)^{n-1}\lambda_{n,1}}{3\sqrt{2}\mu_{s_{1^\Minus}}\mu_{t^z_{2^\Minus}}}\left(C_{s_{{1^\Minus},2}}C_{t^z_{{ 2^\Minus},1}}-2C_{s_{{1^\Minus},3}}C_{t^z_{{ 2^\Minus},2}}\right),\\
 & A^n_{3}= \frac{\left(-1\right)^{n-1}\lambda_{n,3}}{2\sqrt{6}\mu_{s_{1^\Minus}}\mu_{t^z_{3^\Minus}}}\left(C_{s_{{1^\Minus},2}}C_{t^z_{{ 3^\Minus},1}}-2C_{s_{{1^\Minus},3}}C_{t^z_{{ 3^\Minus},2}}\right),\\
 & B^n_1=0, 
 B^n_2\!=\!\frac{\lambda_{n,3}}{2\sqrt{2}\mu_{t^z_{2^\Minus}}}C_{t^z_{{ 2^\Minus},1}}, 
 B^n_3\!=\!-\frac{\lambda_{n,1}}{\sqrt{6}\mu_{t^z_{3^\Minus}}}C_{t^z_{{ 3^\Minus},1}},
 D^n_{11}\!=\!\frac{1}{6},\\
 & D^n_{22}\!=\!\frac{1}{6\mu^2_{t^x_{ 2^\Minus}}}\!\!\left(\!\lambda_{n,11}(C_{t^x_{{ 2^\Minus},1}}C_{t^x_{{ 2^\Minus},1}}\!\!+\!
C_{t^x_{{ 2^\Minus},2}}C_{t^x_{{ 2^\Minus},2}})\!-\!\lambda_{n,10}C_{t^x_{{ 2^\Minus},3}}C_{t^x_{{ 2^\Minus},3}}\!\right)\!,\\
& D^n_{33}\!=\! \frac{1}{2\mu^2_{t^x_{ 3^\Minus}}}\!\!\left(\!\frac{\lambda_{n,2}}{2}(C_{t^x_{{ 3^\Minus},1}}C_{t^x_{{ 3^\Minus},1}}\!\!+\!
 C_{t^x_{{ 3^\Minus},2}}C_{t^x_{{ 3^\Minus},2}})\!+\!\lambda_{n,14}C_{t^x_{{ 3^\Minus},3}}C_{t^x_{{ 3^\Minus},3}}\!\right)\!,\\
 &D^n_{12}\!=\!\frac{\lambda_{n,1}}{3\sqrt{2}\mu_{t^x_{1^\Minus}}\mu_{t^x_{2^\Minus}}}\!\left(C_{t^x_{{1^\Minus},1}}C_{t^x_{{ 2^\Minus},1}}\!\!+
 C_{t^x_{{1^\Minus},2}}C_{t^x_{{ 2^\Minus},2}}\!\!-2C_{t^x_{{1^\Minus},3}}C_{t^x_{{ 2^\Minus},3}}\right)\!,\\
 &D^n_{13}\!=\!\frac{\lambda_{n,3}}{2\sqrt{6}\mu_{t^x_{1^\Minus}}\mu_{t^x_{3^\Minus}}}\!\left(C_{t^x_{{1^\Minus},1}}C_{t^x_{{ 3^\Minus},1}}\!\!+
 C_{t^x_{{1^\Minus},2}}C_{t^x_{{ 3^\Minus},2}}\!\!-2C_{t^x_{{1^\Minus},3}}C_{t^x_{{ 3^\Minus},3}}\right)\!,\\
 & D^n_{23}\!=\!-\frac{\lambda_{n,3}}{4\sqrt{3}\mu_{t^x_{2^\Minus}}\mu_{t^x_{3^\Minus}}}\!\left(C_{t^x_{{ 2^\Minus},1}}C_{t^x_{{ 3^\Minus},1}}\!\!\!+
 C_{t^x_{{ 2^\Minus},2}}C_{t^x_{{ 3^\Minus},2}}\!\!\!-2C_{t^x_{{ 2^\Minus},3}}C_{t^x_{{ 3^\Minus},3}}\!\right)\!.
 \end{aligned}
 \end{equation}
 
 \normalsize
\section{ DETAILS OF MEAN-FIELD APPROXIMATION}
  \label{mean}
Explicit forms of $H_{nm}$ terms of Eq \ref{ham2}
in the momentum space are given here. 
  \begin{equation}
  \begin{aligned}
  H_{30}\!&=\frac{\epsilon^{\alpha\beta\gamma}}{\sqrt{N^\prime}}\!\!\!\sum\limits_{\bold{p},\bold{k},\eta,\xi,\iota}\!\!\!\! \!Z^{\eta\xi\iota}_{\bold{p}-\bold{k}} \,t^\dagger_{\eta^\Minus,\bold{k}-\bold{p},\alpha}t^\dagger_{\xi^\Minus,\bold{p},\beta}t_{\iota^\Minus,\bold{k},\gamma}+\rm{H.c.}, \\
   H_{40}\!&=\frac{\epsilon^{\alpha\beta\gamma}\epsilon^{\alpha\lambda\nu}}{N^\prime}\sum\limits_{\bold{p},\bold{q},\bold{k},\eta,\xi,\iota,\zeta}M^{\eta\xi\iota\zeta}_{\bold{k}}t^\dagger_{\eta^\Minus,{\bold{p}}+{\bold{k}},\beta}t^\dagger_{\xi^\Minus,{\bold{q}}-{\bold{k}},\lambda}t_{\iota^\Minus,{\bold{q}},\nu}t_{\zeta^\Minus,{\bold{p}},\gamma},\\
   H_{21}\!&=\frac{1}{\sqrt{N^\prime}}\sum\limits_{\bold{p},\bold{k},\eta,\xi}\left[W^{\eta\xi}_{\bold{p}}t^\dagger_{\eta^\Minus,{\bold{k}}-{\bold{p}},\alpha}t^\dagger_{\xi^\Minus,\bold{p},\beta}s_{m,\bold{k}}+\rm{H.c.}
   + W^{\xi\eta}_{\bold{p}}s^\dagger_{m,{\bold{k}}-{\bold{p}}}t^\dagger_{\eta^\Minus,\bold{p},\alpha}t^\dagger_{\xi^\Minus,\bold{k},\alpha}+\rm{H.c.}\right],\\
   H_{22}&=\frac{1}{N^\prime}\sum\limits_{\bold{p},\bold{q},\bold{k},\eta,\xi}\left[N^{\eta\xi}_{\bold{k}}s^\dagger_{m,{\bold{q}}+{\bold{k}}}s_{m,{\bold{p}}+{\bold{k}}}t^\dagger_{\eta^\Minus,\bold{p},\alpha}t_{\xi^\Minus,\bold{q},\alpha}
   +\frac{1}{2}N^{\eta\xi}_{\bold{k}}s^\dagger_{m,{\bold{q}}+{\bold{k}}}s^\dagger_{m,{\bold{p}}-{\bold{k}}}t_{\eta^\Minus,\bold{p},\alpha}t_{\xi^\Minus,\bold{q},\alpha}+\rm{H.c.}\right],
     \end{aligned}
 \end{equation}
 
   \begin{equation}
  \begin{aligned}
  Z^{\eta\xi\iota}_{\bold{k}}&=-i\bar s \sum\limits_{n}\left[g^{\eta\xi\iota}_Z(n)e^{-i\bold{k}\cdot \boldsymbol{n}}+g^{\xi\iota\eta}_{\bar Z}(n)e^{i\bold{k}\cdot \boldsymbol{n}}\right],\\
  M^{\eta\xi\iota\zeta}_{\bold{k}}&=-\frac{1}{2}  \sum\limits_{n}\left[g^{\eta\xi\iota\zeta}_M(n)e^{i\bold{k}\cdot \boldsymbol{n}}+g^{\iota\zeta\eta\xi}_M(n)e^{-i\bold{k}\cdot \boldsymbol{n}}\right],\\
  W^{\eta\xi}_{\bold{k}}&=\bar s  \sum\limits_{n}\left[g^{\eta\xi}_W(n)e^{-i\bold{k}\cdot \boldsymbol{n}}+g^{\xi\eta}_{\bar W}(n)e^{i\bold{k}\cdot \boldsymbol{n}}\right],\\
   N^{\eta\xi}_{\bold{k}}&= \sum\limits_{n}\left[g^{\eta\xi}_N(n)e^{-i\bold{k}\cdot \boldsymbol{n}}+g^{\xi\eta}_N(n)e^{i\bold{k}\cdot \boldsymbol{n}}\right].
   \end{aligned}
 \end{equation}
  The $g$ coefficients are given by
  \begin{equation}
  \begin{aligned}
  g^{\eta\xi\iota}_{ Z}(n)&=J_1\left(A^1_\eta D^4_{\xi\iota}\delta_{n,1}+A^2_\eta D^5_{\xi\iota} \delta_{n,1+2}+A^3_\eta D^6_{\xi\iota} \delta_{n,2}\right)
  +J_2\bigg((A^1_\eta D^5_{\xi\iota} +A^1_\eta D^3_{\xi\iota} +A^2_\eta D^4_{\xi\iota}+ A^6_\eta D^4_{\xi\iota})\delta_{n,1} \\
  &+\left(A^2_\eta D^6_{\xi\iota}+A^3_\eta D^1_{\xi\iota}+A^3_\eta D^5_{\xi\iota}+A^4_\eta D^6_{\xi\iota}\right)\delta_{n,2} 
  + (A^1_\eta D^5_{\xi\iota}+A^2_\eta D^6_{\xi\iota}+A^2_\eta D^4_{\xi\iota}+A^3_\eta D^5_{\xi\iota})\delta_{n,1+2}\bigg),\\
  g^{\eta\xi\iota}_{\bar Z}(n)&=J_1\left(D^1_{\eta\xi} A^4_\iota\delta_{n,1}+D^2_{\eta\xi} A^5_\iota \delta_{n,1+2}+D^3_{\eta\xi} A^6_\iota \delta_{n,2}\right)
  +J_2\bigg((D^1_{\eta\xi} A^5_\iota +D^1_{\eta\xi} A^3_\iota +D^2_{\eta\xi} A^4_\iota+ D^6_{\eta\xi} A^4_\iota)\delta_{n,1} \\
  &+\left(D^2_{\eta\xi} A^6_\iota+D^3_{\eta\xi} A^1_\iota+D^3_{\eta\xi} A^5_\iota+D^4_{\eta\xi} A^6_\iota\right)\delta_{n,2} 
  + (D^1_{\eta\xi} A^5_\iota+D^2_{\eta\xi} A^6_\iota+D^2_{\eta\xi} A^4_\iota+D^3_{\eta\xi} A^5_\iota)\delta_{n,1+2}\bigg),\\
   g^{\eta\xi\iota\zeta}_M(n)&=J_1\left(D^1_{\eta\xi} D^4_{\iota\zeta}\delta_{n,1}+D^2_{\eta\xi} D^5_{\iota\zeta} \delta_{n,1+2}+D^3_{\eta\xi} D^6_{\iota\zeta} \delta_{n,2}\right)
  +J_2\bigg((D^1_{\eta\xi} D^5_{\iota\zeta} +D^1_{\eta\xi} D^3_{\iota\zeta} +D^2_{\eta\xi} D^4_{\iota\zeta}+ D^6_{\eta\xi} D^4_{\iota\zeta})\delta_{n,1} \\
  &+\left(D^2_{\eta\xi} D^6_{\iota\zeta}+D^3_{\eta\xi} D^1_{\iota\zeta}+D^3_{\eta\xi} D^5_{\iota\zeta}+D^4_{\eta\xi} D^6_{\iota\zeta}\right)\delta_{n,2} 
  + (D^1_{\eta\xi} D^5_{\iota\zeta}+D^2_{\eta\xi} D^6_{\iota\zeta}+D^2_{\eta\xi} D^4_{\iota\zeta}+D^3_{\eta\xi} D^5_{\iota\zeta})\delta_{n,1+2}\bigg),\\
  g^{\eta\xi}_W(n)&=J_1\left(B^1_\eta A^4_\xi\delta_{n,1}+B^2_\eta A^5_\xi \delta_{n,1+2}+B^3_\eta A^6_\xi \delta_{n,2}\right)
  +J_2\bigg((B^1_\eta A^5_\xi +B^1_\eta A^3_\xi +B^2_\eta A^4_\xi+ B^6_\eta A^4_\xi)\delta_{n,1} \\
  &+\left(B^2_\eta A^6_\xi+B^3_\eta A^1_\xi+B^3_\eta A^5_\xi+B^4_\eta A^6_\xi\right)\delta_{n,2} 
  + (B^1_\eta A^5_\xi+B^2_\eta A^6_\xi+B^2_\eta A^4_\xi+B^3_\eta A^5_\xi)\delta_{n,1+2}\bigg),\\
   g^{\eta\xi}_{\bar W}(n)&=J_1\left(A^1_\eta B^4_\xi\delta_{n,1}+A^2_\eta B^5_\xi \delta_{n,1+2}+A^3_\eta B^6_\xi \delta_{n,2}\right)
  +J_2\bigg((A^1_\eta B^5_\xi +A^1_\eta B^3_\xi +A^2_\eta B^4_\xi+ A^6_\eta B^4_\xi)\delta_{n,1} \\
  &+\left(A^2_\eta B^6_\xi+A^3_\eta B^1_\xi+A^3_\eta B^5_\xi+A^4_\eta B^6_\xi\right)\delta_{n,2} 
  + (A^1_\eta B^5_\xi+A^2_\eta B^6_\xi+A^2_\eta B^4_\xi+A^3_\eta B^5_\xi)\delta_{n,1+2}\bigg),\\
  g^{\eta\xi}_N(n)&=J_1\left(B^1_\eta B^4_\xi\delta_{n,1}+B^2_\eta B^5_\xi \delta_{n,1+2}+B^3_\eta B^6_\xi \delta_{n,2}\right)
  +J_2\bigg((B^1_\eta B^5_\xi +B^1_\eta B^3_\xi +B^2_\eta B^4_\xi+ B^6_\eta B^4_\xi)\delta_{n,1} \\
  &+\left(B^2_\eta B^6_\xi+B^3_\eta B^1_\xi+B^3_\eta B^5_\xi+B^4_\eta B^6_\xi\right)\delta_{n,2} 
  + (B^1_\eta B^5_\xi+B^2_\eta B^6_\xi+B^2_\eta B^4_\xi+B^3_\eta B^5_\xi)\delta_{n,1+2}\bigg).
   \end{aligned}
  \end{equation}
  
  Here, $\eta,\xi, \iota,\zeta =1,2,3$ and $\alpha,\beta,\gamma=x,y,z$.  
$m=2$ and $1^\Minus$ for the regions R$_1$ and R$_2$, respectively.  
  
The coefficients, $ X^{\eta\xi}_{\bold{k}}$ and $ Y^{\eta\xi}_{\bold{k}}$ are given as
     \begin{equation}
  \begin{aligned}
   X^{\eta\xi}_{\bold{k}}&=\left(E_{t_\eta}-\mu\right)\left(\delta_{\eta,1^\Minus}\delta_{\xi,1^\Minus}+\delta_{\eta,2^\Minus}\delta_{\xi,2^\Minus}
   +\delta_{\eta,3^\Minus}\delta_{\xi,3^\Minus}\right)+  Y^{\eta\xi}_{\bold{k}},\\
    Y^{\eta\xi}_{\bold{k}}&=\bar s^2\sum\limits_{n}2g^{\eta\xi}(n)\cos\left(\bold{k}\cdot \boldsymbol{n}\right), \quad {\rm where}\\
  g^{\eta\xi}(n)&=J_1\left(A^1_\eta A^4_\xi\delta_{n,1}+A^2_\eta A^5_\xi \delta_{n,1+2}+A^3_\eta A^6_\xi \delta_{n,2}\right)
  +J_2\bigg((A^1_\eta A^5_\xi +A^1_\eta A^3_\xi +A^2_\eta A^4_\xi+ A^6_\eta A^4_\xi)\delta_{n,1} \\
  &+\left(A^2_\eta A^6_\xi+A^3_\eta A^1_\xi+A^3_\eta A^5_\xi+A^4_\eta A^6_\xi\right)\delta_{n,2} 
  + (A^1_\eta A^5_\xi+A^2_\eta A^6_\xi+A^2_\eta A^4_\xi+A^3_\eta A^5_\xi)\delta_{n,1+2}\bigg). 
   \end{aligned}
  \end{equation}
  \normalsize
Here, $\eta,\xi=1,2,3$ and $E_{t_\eta}$ is the triplet energy of the single 
plaquette. $n = 1,2$ correspond to the 
NN vectors $\boldsymbol{\tau}_1$ and $\boldsymbol{\tau}_2$. The expressions of all $g$ coefficients in the region R$_2$ will be same with the interchange of $A$ and $B$. 

The analytic procedure used to diagonalize the mean-field 
Hamiltonian (Eq \ref{ham4}) expressed in terms of bosonic operators 
has been described below.
Instead of ${H_{\bold{k}}}$, ${ I_B}{H_{\bold{k}}}$ has been 
diagonalized \cite{Colpa}, where ${I_B}$=diag[$1,1,1,-1,-1,-1$]. 
The characteristic equation and positive eigenvalues of the matrix ${ I_B}{H_{\bold{k}}}$ are 
written below. 
 \begin{equation}
  \begin{aligned}
& \Omega_\bold{k}^6+ a_{2,\bold{k}}\Omega_\bold{k}^4+a_{1,\bold{k}}\Omega_\bold{k}^2+a_{0,\bold{k}}=0, \\
& \Omega_{\eta,\bold{k}}=\left[2\sqrt{-Q_\bold{k}}\cos(\frac{\theta}{3}-\frac{2\pi p}{3})-\frac{a_{2,\bold{k}}}{3}\right]^\frac{1}{2}, \quad 
 Q_\bold{k}=\frac{3a_{1,\bold{k}}-a^2_{1,\bold{k}}}{9}, \\
& R_\bold{k}=\frac{9a_{2,\bold{k}}a_{1,\bold{k}}-27a_{0,\bold{k}}-2a^3_{1,\bold{k}}}{54}, \quad
 \cos(\theta)=\frac{-R_\bold{k}}{Q_\bold{k}\sqrt{-Q_\bold{k}}}, 
 \end{aligned}
\end{equation}
where, $p=0,1,2$ and $\eta=1,2,3$.
The coefficients, $a_{i,\bold{k}}$ are given below. 
 \begin{equation}
  \begin{aligned}
   a_{2,\bold{k}}&=-\left(w^2_{11,\bold{k}}+w^2_{22,\bold{k}}+w^2_{33,\bold{k}}\right), \\
   a_{1,\bold{k}}&=w^2_{11,\bold{k}}w^2_{22,\bold{k}}+w^2_{11,\bold{k}}w^2_{33,\bold{k}}+w^2_{22,\bold{k}}w^2_{33,\bold{k}}
 -4(Y^{12}_\bold{k})^2\left(X^{11}_\bold{k}-Y^{11}_\bold{k}\right)\left(X^{22}_\bold{k}-Y^{22}_\bold{k}\right)\\
& -4(Y^{23}_\bold{k})^2\left(X^{22}_\bold{k}-Y^{22}_\bold{k}\right)\left(X^{33}_\bold{k}-Y^{33}_\bold{k}\right)
 -4(Y^{13}_\bold{k})^2\left(X^{11}_\bold{k}-Y^{11}_\bold{k}\right)\left(X^{33}_\bold{k}-Y^{33}_\bold{k}\right), \\
  a_{0,\bold{k}}&=\left(X^{11}_\bold{k}-Y^{11}_\bold{k}\right)\left(X^{22}_\bold{k}-Y^{22}_\bold{k}\right)\left(X^{33}_\bold{k}-Y^{33}_\bold{k}\right)
  \big[ 4(Y^{12}_\bold{k})^2\left(X^{33}_\bold{k}+Y^{33}_\bold{k}\right)
 + 4(Y^{23}_\bold{k})^2\left(X^{11}_\bold{k}+Y^{11}_\bold{k}\right) \\
   &+ 4(Y^{13}_\bold{k})^2\left(X^{22}_\bold{k}+Y^{22}_\bold{k}\right)-16Y^{12}_\bold{k}Y^{13}_\bold{k}Y^{23}_\bold{k}
 -\left(X^{11}_\bold{k}+Y^{11}_\bold{k}\right)\left(X^{22}_\bold{k}+Y^{22}_\bold{k}\right)\left(X^{33}_\bold{k}+Y^{33}_\bold{k}\right)\big], 
   \end{aligned}
\end{equation}
where, $w^2_{\eta\xi,\bold{k}}=(X^{\eta\xi}_\bold{k})^2-(Y^{\eta\xi}_\bold{k})^2$ with $\eta,\xi=1,2,3$.

Using the procedure developed before \cite{Colpa}, analytic expressions of 
the Bogoliubov coefficients have been obtained.
The Bogoliubov coefficients $u^{\eta\xi}_\bold{k}$ and $v^{\eta\xi}_\bold{k}$ are  
 \begin{equation}
  \begin{aligned}
  u^{\eta\xi}_\bold{k}=\frac{\phi_\bold{k}^{\eta\xi}+\psi_\bold{k}^{\eta\xi}}{2}, \quad 
  v^{\eta\xi}_\bold{k}=\frac{\phi_\bold{k}^{\eta\xi}-\psi_\bold{k}^{\eta\xi}}{2},
  \end{aligned}
\end{equation}
with $\eta,\xi=1,2$  and  3. Where,
 \begin{equation}
  \begin{aligned}
\phi_\bold{k}^{1\eta}&=x_{\eta,\bold{k}}\sqrt{X_\bold{k}^{11}-Y_\bold{k}^{11}}, \quad   
\phi_\bold{k}^{2\eta}=y_{\eta,\bold{k}}\sqrt{X_\bold{k}^{22}-Y_\bold{k}^{22}},\quad  
\phi_\bold{k}^{3\eta}= z_{\eta,\bold{k}} \sqrt{X_\bold{k}^{33}-Y_\bold{k}^{33}},\\
\psi_\bold{k}^{1\eta}&=\big(x_{\eta,\bold{k}}(X_\bold{k}^{11}+Y_\bold{k}^{11})\sqrt{X_\bold{k}^{11}-Y_\bold{k}^{11}}  
+2y_{\eta,\bold{k}} Y_\bold{k}^{12}\sqrt{X_\bold{k}^{22}-Y_\bold{k}^{22}} 
+2z_{\eta,\bold{k}} Y_\bold{k}^{13}\sqrt{X_\bold{k}^{33}-Y_\bold{k}^{33}}\big)  /\Omega_{\eta,\bold{k}}, \\
\psi_\bold{k}^{2\eta}&=\big(2 x_{\eta,\bold{k}} Y_\bold{k}^{12}\sqrt{X_\bold{k}^{11}-Y_\bold{k}^{11}}
+y_{\eta,\bold{k}} (X_\bold{k}^{22}+Y_\bold{k}^{22})\sqrt{X_\bold{k}^{22}-Y_\bold{k}^{22}} 
+2z_{\eta,\bold{k}} Y_\bold{k}^{23}\sqrt{X_\bold{k}^{33}-Y_\bold{k}^{33}}\big)/\Omega_{\eta,\bold{k}},\\
\psi_\bold{k}^{3\eta}&=\big(2 x_{\eta,\bold{k}} Y_\bold{k}^{13}\sqrt{X_\bold{k}^{11}-Y_\bold{k}^{11}}
+2y_{\eta,\bold{k}} Y_\bold{k}^{23}\sqrt{X_\bold{k}^{22}-Y_\bold{k}^{22}} 
+z_{\eta,\bold{k}} (X_\bold{k}^{33}+Y_\bold{k}^{33})\sqrt{X_\bold{k}^{33}-Y_\bold{k}^{33}}\big)/\Omega_{\eta,\bold{k}},\\
x_{\eta,\bold{k}}&=\frac{M_{\eta,\bold{k}}}{\sqrt{G_{\eta,\bold{k}}}}, \quad
y_{\eta,\bold{k}}=\frac{1}{\sqrt{G_{\eta,\bold{k}}}},\quad
z_{\eta,\bold{k}}=\frac{N_{\eta,\bold{k}}}{\sqrt{G_{\eta,\bold{k}}}},\quad
M_{\eta,\bold{k}}=\frac{A_{\bold{k}}C_{\bold{k}}-(w^2_{22,\bold{k}}-\Omega^2_{\eta,\bold{k}})B_{\bold{k}}}
{A_{\bold{k}}B_{\bold{k}}-(w^2_{11,\bold{k}}-\Omega^2_{\eta,\bold{k}})C_{\bold{k}}}, \\
N_{\eta,\bold{k}}&=\frac{C_{\bold{k}}+B_{\bold{k}}M_{\eta,\bold{k}}}{(\Omega^2_{\eta,\bold{k}}-w^2_{33,\bold{k}})},\quad
G_{\eta,\bold{k}}=\Omega_{\eta,\bold{k}}\left[1+M^2_{\eta,\bold{k}}+N^2_{\eta,\bold{k}}\right],\\
A_{\bold{k}}&=2 Y_\bold{k}^{12}\sqrt{X_\bold{k}^{11}-Y_\bold{k}^{11}}\sqrt{X_\bold{k}^{22}-Y_\bold{k}^{22}},  \\
B_{\bold{k}}&=2 Y_\bold{k}^{13}\sqrt{X_\bold{k}^{11}-Y_\bold{k}^{11}}\sqrt{X_\bold{k}^{33}-Y_\bold{k}^{33}}, \\
C_{\bold{k}}&=2 Y_\bold{k}^{23}\sqrt{X_\bold{k}^{22}-Y_\bold{k}^{22}}\sqrt{X_\bold{k}^{33}-Y_\bold{k}^{33}}. 
    \end{aligned}
\end{equation}
\end{widetext}
  \end{document}